\documentclass{aa}
\usepackage{subcaption}
\usepackage{graphicx}
\usepackage{txfonts}
\usepackage{natbib}
\usepackage{hyperref}
\bibpunct{(}{)}{;}{a}{}{,} 

\begin{document}

\title{Particle acceleration with anomalous pitch angle scattering in 2D MHD reconnection simulations}
\titlerunning{Particle acceleration and pitch angle scattering in 2D reconnection simulations}
\author{A. Borissov
        \inst{1}
        \and
        E.P. Kontar
        \inst{2}
        \and
        J. Threlfall
        \inst{1}
        \and
        T. Neukirch
        \inst{1}}
\institute{School of Mathematics and Statistics, University of St Andrews, St Andrews KY16 9SS, U.K.
\and
School of Physics and Astronomy, University of Glasgow, Glasgow G12 8QQ, U.K.}


\abstract{The conversion of magnetic energy into other forms (such as plasma heating, bulk plasma flows, and non-thermal particles) during solar flares is one of the outstanding open problems in solar physics. It is generally accepted that magnetic reconnection plays a crucial role in these conversion processes. In order to achieve the rapid energy release required in solar flares, an anomalous resistivity, which is orders of magnitude higher than the Spitzer resistivity, is often used in magnetohydrodynamic (MHD) simulations of reconnection in the corona. The origin of Spitzer resistivity is based on Coulomb scattering, which becomes negligible at the high energies achieved by accelerated particles. As a result, simulations of particle acceleration in reconnection events are often performed in the absence of any interaction between accelerated particles and any background plasma. This need not be the case for scattering associated with anomalous resistivity caused by turbulence within solar flares, as the higher resistivity implies an elevated scattering rate. We present results of test particle calculations, with and without pitch angle scattering, subject to fields derived from MHD simulations of two-dimensional (2D) X-point reconnection. Scattering rates proportional to the ratio of the anomalous resistivity to the local Spitzer resistivity, as well as at fixed values, are considered. Pitch angle scattering, which is independent of the anomalous resistivity, causes higher maximum energies in comparison to those obtained without scattering. Scattering rates which are dependent on the local anomalous resistivity tend to produce fewer highly energised particles due to weaker scattering in the separatrices, even though scattering in the current sheet may be stronger when compared to resistivity-independent scattering. Strong scattering also causes an increase in the number of particles exiting the computational box in the reconnection outflow region, as opposed to along the separatrices as is the case in the absence of scattering.}


\maketitle

\section{Introduction}

Solar flare energy release is commonly attributed to magnetic reconnection, during which magnetic energy is converted into other forms, such as plasma heating, bulk plasma flows, and non-thermal accelerated particles \citep[for reviews of solar flare observations and theory see e.g.][]{priest-forbes2002,benz2008,shibata-magara2011}. Despite many years of research, the physics behind these processes is still not entirely understood. Fast magnetic reconnection is fundamentally based upon particle scattering \citep{treumann-baumjohann2015}, which causes a restructuring of the magnetic field through diffusion of the magnetic field with respect to the plasma. With any scattering model there will be an associated resistivity. In the case of binary Coulomb collisions the associated resistivity is the Spitzer resistivity, which is typically too small to account for the high rate of energy release in solar flares \citep{birn-priest2007}. 

The introduction of anomalous resistivity, caused by turbulent processes, could account for the rate of energy release during flares \citep[for discussions on the origin of anomalous resitivity see e.g.][]{papadopoulos1977,treumann2001}. In addition, multiple flare models require the presence of particle scattering due to turbulence \citep[see e.g.][]{petrosian2012}, and there has been evidence for the presence of magnetohydrodynamic (MHD) turbulence in solar flares \citep[e.g.][]{kontar-et-al2017}. Vlasov or particle-in-cell simulations are required in order to capture the physics of turbulent processes in magnetic reconnection. Unfortunately these simulations are too computationally expensive to model the whole of a solar flare, so an MHD approach is often used. While MHD allows the simulation of larger lengthscales and timescales, it cannot capture the microscopic physics involved in collisionless reconnection and hence requires the specification of an anomalous resistivity affecting the electromagnetic field evolution through the magnetic induction equation and Ohm's law. In general, for non-zero resistivity a component of the electric field will be directed parallel to the magnetic field, which will result in acceleration of non-thermal particles. 

Acceleration due to parallel electric field is one of the main acceleration mechanisms thought to produce a non-thermal particle population which is the cause of the observed hard X-ray radiation in solar flares \citep[for reviews of particle acceleration mechanisms see][]{zharkova-et-al2011,cargill-et-al2012}. Test particle simulations of acceleration in MHD simulations of magnetic reconnection in two dimensions \citep[see e.g. ][]{gordovskyy-et-al2010a}, and in various scenarios in three dimensions \citep[e.g.][]{gordovskyy-et-al2013,gordovskyy-et-al2014,threlfall-et-al2016a}, have been performed, both with and without Coulomb scattering. In all cases, however, an anomalous resistivity was specified in the MHD simulation. In order to have a more consistent description of the interaction between the accelerated particles and the background an enhanced anomalous scattering rate (relative to Coulomb scattering) should be used. One possibility is the use of pitch angle scattering \citep[see e.g. ][]{jeffrey-et-al2014,kontar-et-al2014,bian-et-al2016,bian-et-al2017}, with a scattering rate that is dependent on the resistivity.

In this paper we complement previous work by presenting the results of test particle simulations including pitch angle scattering in fields generated by two-dimensional (2D) MHD simulations. We examine the impact of pitch angle scattering, with varying dependencies of the scattering rate on the velocity of the particle and anomalous resistivity used in the MHD simulations. Individual trajectories as well as energy spectra and spatial distributions are produced. The layout of the remainder of this paper is as follows: in Section \ref{mhd} we describe the configuration and results of 2D MHD reconnection simulations. Section \ref{particle-eqns} describes our modifications to the guiding centre approach to incorporate pitch angle scattering. We present the results of test particle simulations in Section \ref{results} along with conclusions in Section \ref{conclusions}.


\section{MHD simulations}
\label{mhd}


We solve the standard resistive MHD equations \citep[see e.g.][]{priest2014} given by Equations~\ref{mhd1}-\ref{mhd6}:

\begin{eqnarray}
\frac{\partial \rho}{\partial t}&=&- \nabla\cdot(\rho \mathbf{v}) \label{mhd1},\\
\frac{\partial\textbf v}{\partial t} + \textbf v \cdot \nabla \textbf v&=&\frac{1}{\rho}\mathbf{j}\times\mathbf{B}-\frac{1}{\rho}\nabla P \label{mhd2},\\
\frac{\partial \mathbf{B}}{\partial t}&=&-\nabla\times\mathbf{E} \label{mhd3},\\
\frac{\partial \epsilon}{\partial t} + \textbf v \cdot \nabla \epsilon &=&-\frac{P}{\rho}\nabla\cdot\mathbf{v}+\frac{\eta_a}{\rho}j^{2} \label{mhd4},\\
\mathbf{E}+\mathbf{v}\times\mathbf{B}&=&\eta_a \mathbf{j} \label{mhd5},\\
\nabla\times\mathbf{B}&=& \mathbf{j}, \label{mhd6}
\end{eqnarray}
using the \textit{Lare2d} code \citep{arber-et-al2001}, with normalising scales given by $\hat L = 10 \unit m, \hat B = 0.03 \unit T$ and $\hat \rho = 1.67 \times 10^{-12} \unit{kg \cdot m}^{-3}$. The choice of $\hat \rho$ is reflective of the coronal environment \citep{priest2014}, while $\hat B$ is similar to that used in other simulations of magnetic reconnection \citep[e.g.][]{gordovskyy-et-al2010a}. The lengthscale is chosen to be comparable to the current sheet size, which is not well constrained for the solar corona; current sheet sizes similar to ours have been used \citep[see e.g.][]{litvinenko1996,wood-neukirch2005}, however so have much larger ones \citep[e.g.][]{kliem1994,gordovskyy-et-al2010a}. This choice of lengthscale pushes the limits of the applicability of MHD within the solar corona, however it was used in order to achieve a compromise between the use of self-consistent electromagnetic fields (from the MHD simulation) while at the same time incorporating aspects of microscopic physics into the particle acceleration picture (without the use of kinetic simulations). In the absence of scattering, test particle energies scale with the square of length, meaning that orbit calculations performed with a given choice of lengthscale can be extrapolated by simply adjusting the energies appropriately. This is not generally the case with scattering included, as the mean free path associated with the scattering introduces a scale independent of the MHD lengthscale which impacts the particle orbits. Increasing the lengthscale substantially, without changing the scattering mean free path, would result in particle orbit computation becoming prohibitively computationally expensive. Although it would be possible to circumvent this issue by, for example, restricting the domain size within which particle orbit calculations are performed, doing so would restrict the effect of the geometrical configuration of the MHD fields on the particle simulation. The normalising scales for all other parameters come from combinations of $\hat L, \hat B,$ and $\hat \rho$ and are quoted in Table \ref{norms}. We set the anomalous resistivity ($\eta_a$) to zero where the critical current is below a threshold value of $j_{\text{crit}} = 1$, while $\eta_a = 1\times 10^{-4}$ where the current exceeds $j_{\text{crit}}$ (values of $j_{\text{crit}}$ and $\eta_a$ are given in normalised units).

\begin{center}
\begin{table}[ht]
\caption{Normalisation constants for Lare2d. Only the length, magnetic field, and density scaling are specified, while the rest are calculated.}
\label{norms}
\begin{tabular}{cccc}
\hline
Quantity & Normalising value & Quantity & Normalising value \\
\hline
\vspace{2pt}
 $\hat L$ & 10 \unit m & $\hat B$ & $3\times 10^{-2}$ T \\
 $\hat \rho$ & $1.67\times 10^{-12} \unit{kg \cdot m}^{-3}$ & $\hat v$ &  $2.07\times 10^{7} \unit{m \cdot s}^{-1}$\\
 $\hat \varepsilon$ &  $4.28\times 10^{14} \unit{J \cdot kg^{-1}}$ & $\hat t$ & $4.83 \times 10^{-7} \unit s$ \\
 $\hat j$ &  $2.39\times 10^{3} \unit{A \cdot m}^{-2}$& $\hat E$ &  $6.21\times 10^{5} \unit{V \cdot m}^{-1}$\\
 $\hat \eta$ & $260\unit{\Omega \cdot m}$ & $\hat T$ &  $6.23\times 10^{10} \unit{K}$ \\
\hline
\end{tabular}
\end{table}
\end{center}

\begin{figure*}[ht]
\begin{subfigure}[b]{0.49\textwidth}
\includegraphics[width = 0.85\textwidth]{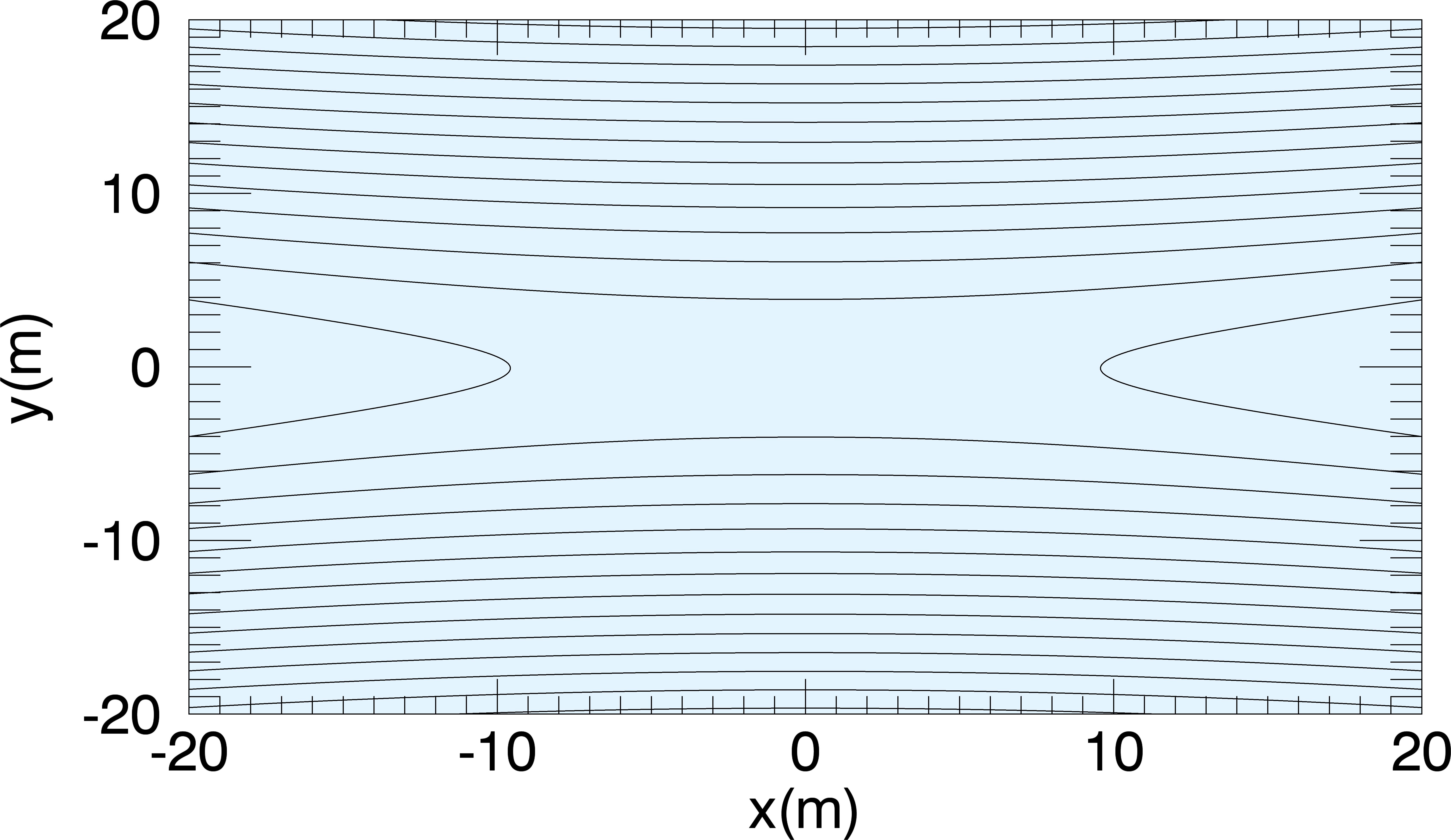}
\subcaption{Initial conditions}\label{mhd_snapshot1}
\end{subfigure}
\begin{subfigure}[b]{0.49\textwidth}
\includegraphics[width = \textwidth]{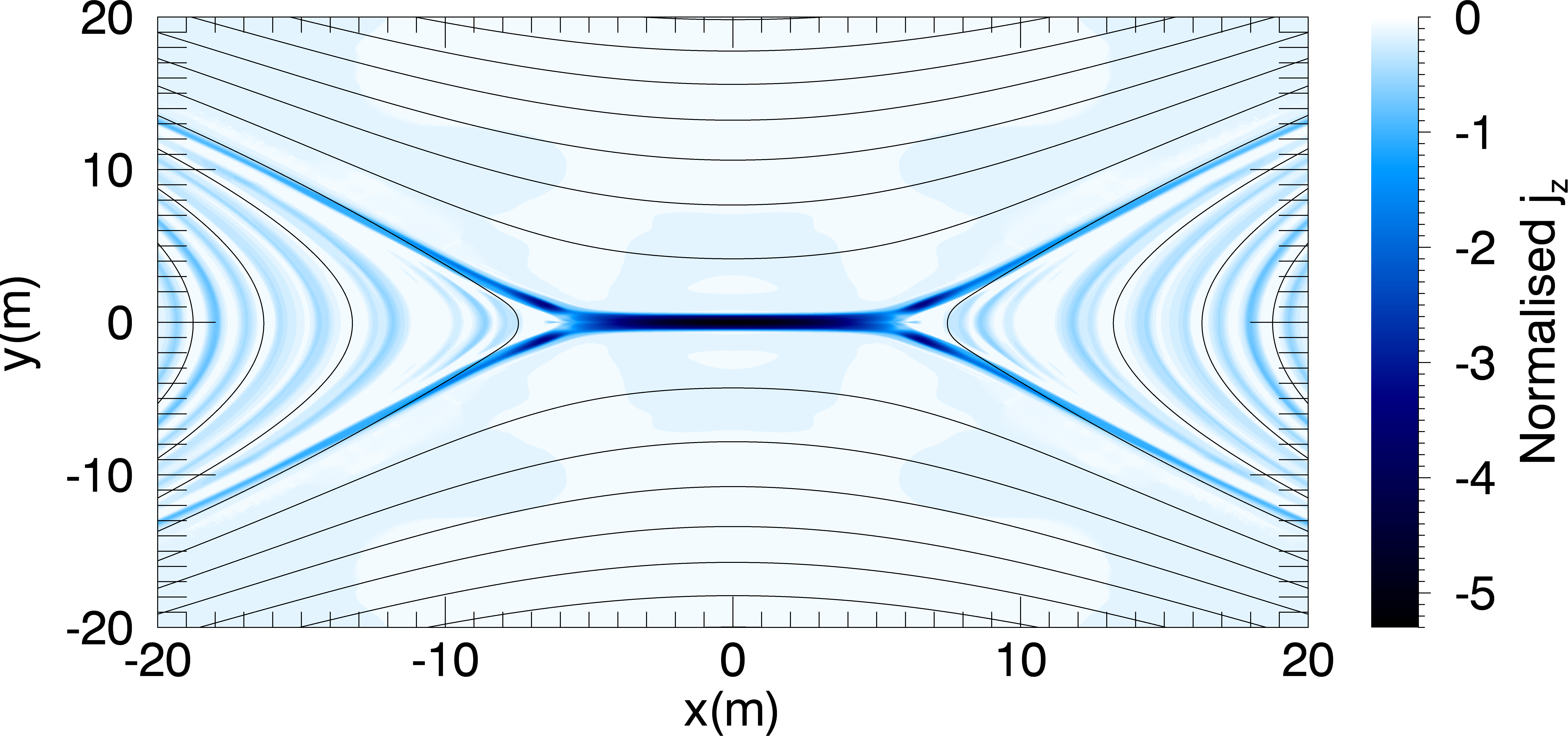}
\subcaption{Snapshot taken from $\eta = 10^{-4}$ simulation}\label{mhd_snapshot2}
\end{subfigure}
\\
\begin{subfigure}[b]{0.49\textwidth}
\includegraphics[width = 0.85\textwidth]{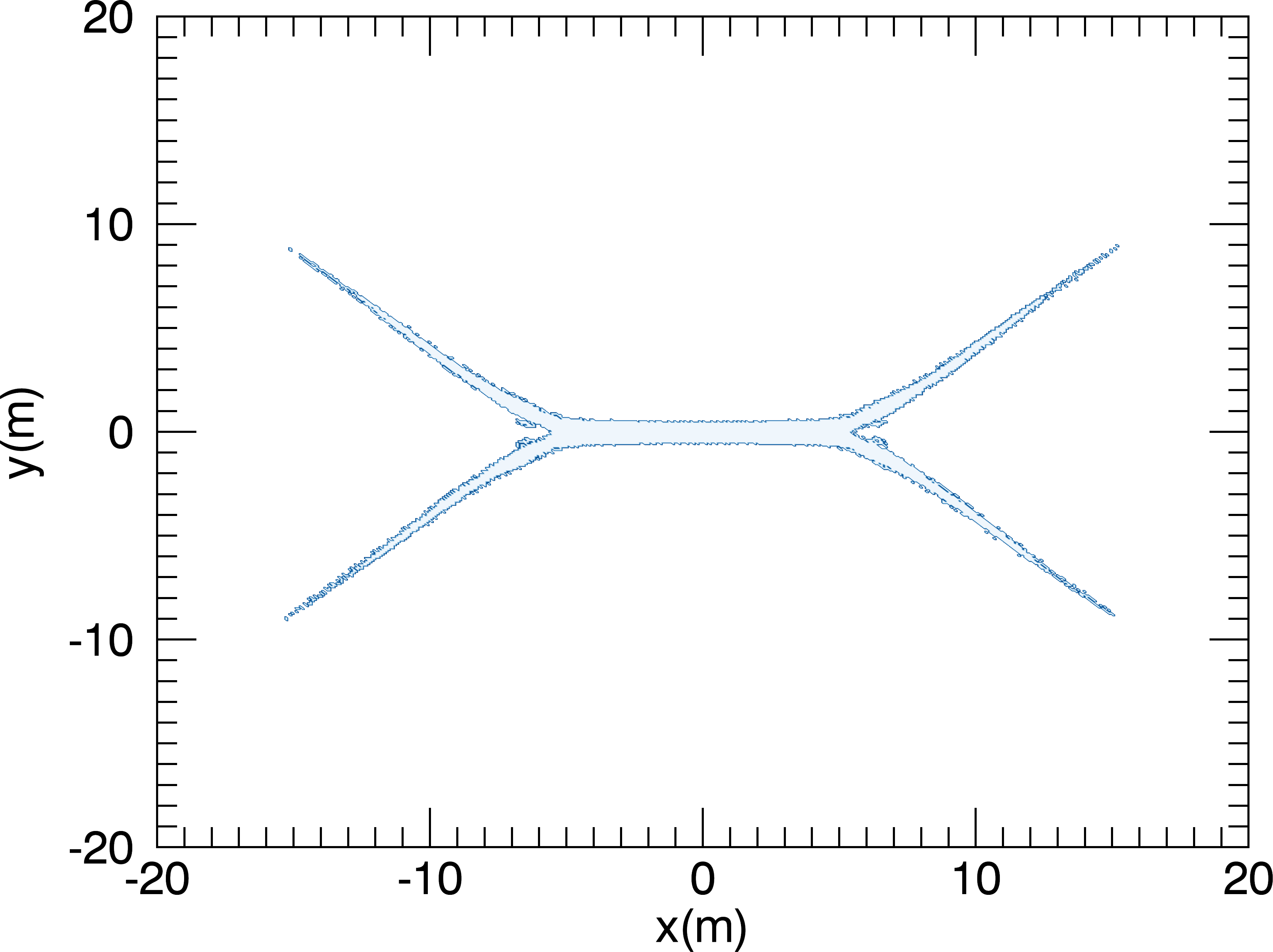}
\subcaption{Contour of non-zero resistivity}\label{mhd_eta}
\end{subfigure}
\begin{subfigure}[b]{0.49\textwidth}
\includegraphics[width = 0.85\textwidth]{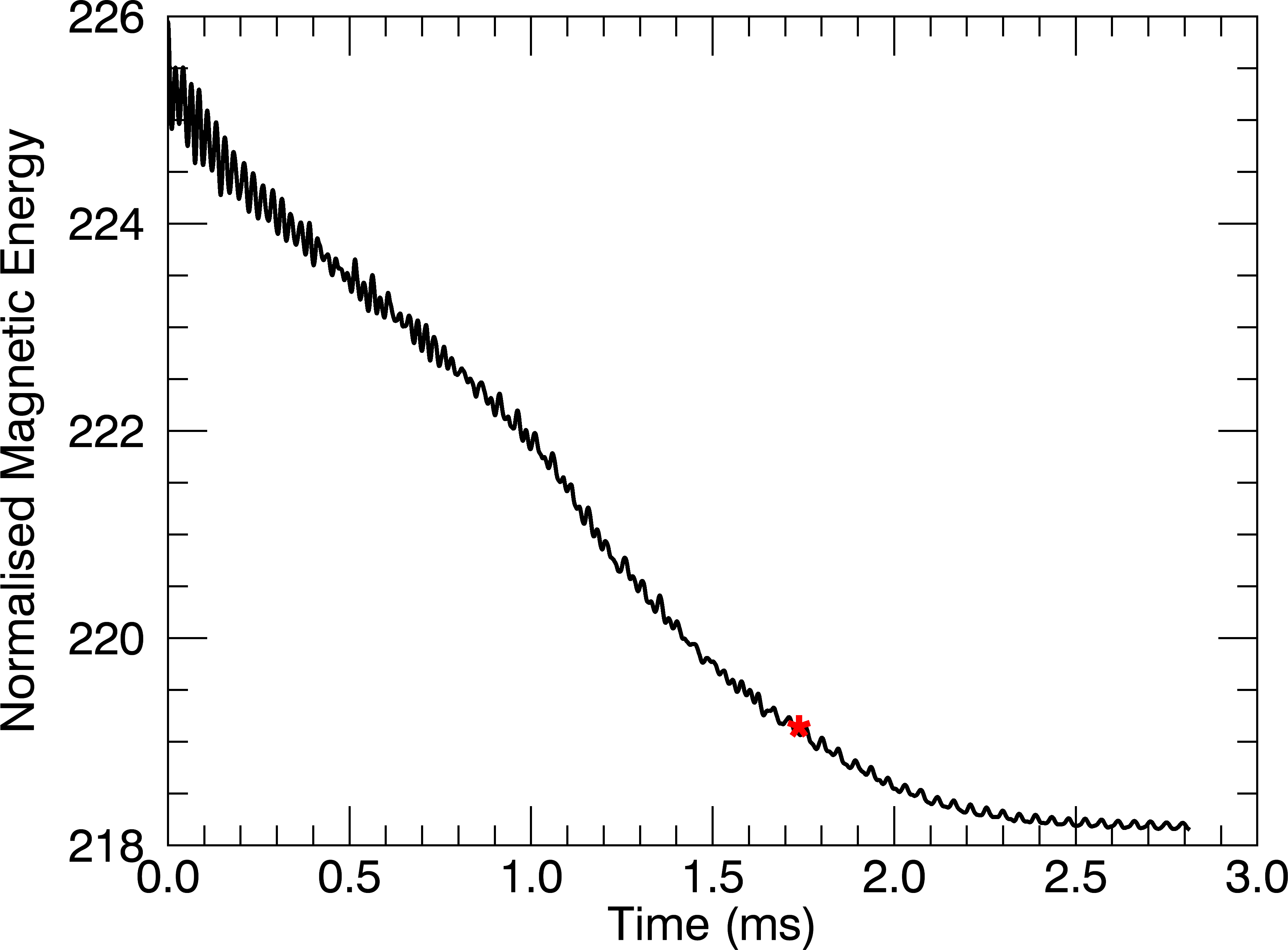}
\subcaption{Time evolution of non-dimensional magnetic energy}\label{mhd_energy}
\end{subfigure}
\caption{Magnetic field lines (black) and out of plane electric field (colour) for \subref{mhd_snapshot1} the initial conditions of the MHD simulation, and \subref{mhd_snapshot2} the chosen snapshot into which test particles are injected. We present only the subset of the MHD simulation domain which is within the test particle computational box. Panel \subref{mhd_eta} shows the areas where current density exceeds the threshold value for triggering anomalous resistivity and coincides with the region where scattering takes place. Panel \subref{mhd_energy} presents the evolution of the magnetic energy in the simulation, with the red star indicating the time at which the snapshot used for the particle simulation is taken.}
\label{mhd_snapshots}
\end{figure*}

Our simulation of 2D magnetic reconnection starts with an isothermal force-free Harris sheet whose magnetic field is perturbed in order to initiate reconnection. The equations specifying the initial conditions for the MHD simulation are given in Equations~\ref{mhdic1}-\ref{mhdic5}:
\begin{equation}\label{mhdic1}
\frac{B_x}{\hat B} = \tanh(y) - \frac{b_1}{b_0}k_y\cos(k_x x)\sin(k_y y),
\end{equation}
\begin{equation}\label{mhdic2}
\frac{B_y}{\hat B} = \frac{b_1}{b_0}k_x\cos(k_y y)\sin(k_x x),
\end{equation}
\begin{equation}\label{mhdic3}
\frac{B_z}{\hat B} = \text{sech}(y),
\end{equation}
\begin{equation}\label{mhdic5}
\frac{\varepsilon}{\hat \varepsilon} = \frac{T_0/\hat T}{m_r(\gamma_p - 1)},
\end{equation}
where $b_1/b_0 = 0.3$, $T_0 = 10^6 \unit K$, $m_r = 1.2$ is the reduced mass for coronal plasma normalised to the proton mass, and $\gamma_p = 5/3$ is the ratio of specific heats. We specify the initial density to be uniform at a value of $5\hat \rho$. Our domain has size 15 in the $x$-direction and 60 in the $y$-direction so that our choices of $k_x = 2\pi/15$, $k_y = 2\pi/60$ ensure the perturbation has one period within the domain in both directions. Periodic boundary conditions in the $x$-direction and closed boundary conditions in the $y$-direction are imposed. The magnetic field corresponding to the initial conditions is shown in Figure \ref{mhd_snapshot1}. 

We evolve the MHD simulation until the reconnection rate drops to near-zero and we use an individual snapshot from the simulation (shown in Figure \ref{mhd_snapshot2}) during the reconnecting phase into which we insert test particles to compute particle orbits. For simplicity we pick a single MHD snapshot as the electromagnetic field structure changes on a longer timescale than the particle evolution. We shall see in Section \ref{spectra} that the majority of the particle orbits' durations are less than 0.1\unit{ms} and the MHD fields do not vary a great deal during the main reconnection phase which lasts approximately 1\unit{ms} (this can be seen from the evolution of the magnetic energy in Figure \ref{mhd_energy}, which steadily decreases between 1 and 2\unit{ms}). 

\section{Governing equations for test particle evolution}
\label{particle-eqns}
Charged particle evolution is governed by the Lorentz force law, $\frac{d\textbf v}{dt} = q\left( \textbf E + \textbf v\times \textbf B \right)$, which can, in principle, be solved numerically for the trajectory of the particle. Unfortunately the timestep required to resolve the evolution of the test particle is too small to be practical (for the magnetic field strengths typical of the corona). A common alternative is to use the guiding centre approximation when computing particle orbits \citep[see for example][]{gordovskyy-et-al2010a,threlfall-et-al2016a,borissov-et-al2016}. In this approach the position of the test particle is averaged over a gyration period (this averaged position is referred to as the guiding centre). This method allows the use of longer timesteps, because the particle gyration need not be resolved temporally. The equations for the evolution of the guiding centre in prescribed electromagnetic fields are given by \cite{northrop1963} and are reproduced in Equations~\ref{NRGC1}-\ref{NRGC2}:

\begin{align}\label{NRGC1}
\dot{\textbf R}_{\perp} &= \frac{\textbf b}{B} \times \left[ -\textbf E + \frac{\mu}{\gamma e} \nabla B + \frac{mU}{e} \frac{d\textbf b}{dt} + \frac{m\gamma}{e}\frac{d\textbf u_{E}}{dt} \right. \nonumber \\ &\hspace{20pt} \left. + \frac{U}{\gamma}E_{\parallel}\textbf u_{E} + \frac{\mu}{\gamma e} \textbf u_{E} \frac{\partial B}{\partial t} \right],
\end{align}

\begin{equation}\label{NRGC2}
m\frac{dU}{dt} = m\gamma \textbf u_{E} \cdot \frac{d\textbf b}{dt} + eE_{\parallel} - \frac{\mu}{\gamma}\frac{\partial B}{\partial s},
\end{equation}
where the Lorentz factor, $\gamma$, is given by:

\begin{equation}\label{NRGC3}
\gamma = \sqrt{1 + \frac{U^2 + u_E^2}{c^2} + \frac{2\mu B}{mc^2}}.
\end{equation}
Here $\textbf R$ denotes the guiding centre position, and $\dot {\textbf R}_\perp$ is the drift velocity of the guiding centre perpendicular to the magnetic field. The E cross B drift of the guiding centre is given by $\textbf u_E = \gamma \textbf V_E = \gamma\textbf E \times \textbf B/B^2$, $U = \gamma v_\parallel = \gamma\textbf v \cdot \textbf b$ is the velocity of the guiding centre parallel to the magnetic field, and $\textbf b$ is the unit vector in the direction of the magnetic field. The quantity $\frac{\partial B}{\partial s}$ is the rate of change of the magnetic field strength along the guiding centre trajectory. Finally, $\mu = m\gamma^2 v_{\perp}^{2}/(2B)$ is the magnetic moment, $v_\perp = v_{tot} \sin\theta$ is the gyrational component of the total particle velocity, $v_{tot} = \left| \textbf v \right|$, $m$ is the electron mass, and $e$ the electron charge. The guiding centre approach is valid as long as the length and timescales on which the underlying fields vary are large compared with the particle gyroradius and gyroperiod. In our simulations the maximum value of the ratio of the electron gyroradius to the width of the current sheet is approximately 0.03, while the maximum value of the ratio of the electron gyroperiod to the MHD timescale is 0.007, justifying our use of the guiding centre model. We use the relativistic version of the guiding centre equations even though the particle energies we obtain are generally non-relativistic.

In regions where $\eta_a = 0$ we solve Equations \ref{NRGC1}-\ref{NRGC3} with an adaptive timestep 4th order Runge Kutta scheme. The guiding centre equations conserve the magnetic moment along the particle trajectory, which cannot be true if pitch angle scattering occurs. By modifying the magnetic moment, along with self-consistently modifying $U$, we can introduce pitch angle scattering into the governing equations of particle motion. To account for pitch angle scattering in regions where $\eta_a\neq 0$, in addition to solving Equations~\ref{NRGC1}-\ref{NRGC2} we also solve:
\begin{equation}\label{sde1}
d\gamma =  \dot \gamma dt,
\end{equation}
\begin{equation}\label{sde2}
d\beta =  (\dot \beta + F_\beta) dt + \sqrt{2D_{\beta\beta}}dW,
\end{equation}
where $\beta = \cos \theta$, and $dW = \zeta \sqrt{dt}$ and $\zeta$ is a normally distributed random variable. Expressions for $\dot \gamma$ and $\dot \beta$ are given by:
\begin{equation}\label{sde3}
\dot \gamma = \frac 1 2 \left( 1 + \frac{U^2}{c^2} + \frac{2\mu B}{mc^2}\right)^{-1/2} \left( \frac{2U}{c^2}\frac{dU}{dt} + \frac{2\mu}{mc^2} \frac{dB}{dt} \right) \left( 1 - \frac{V_E^2}{c^2} \right)^{-1/2},
\end{equation}
\begin{equation}\label{sde4}
\dot \beta = \left( \frac{1}{U}\frac{dU}{dt} - \frac{1}{2B}\frac{dB}{dt} \right) \beta\left( 1-\beta^2 \right).
\end{equation}
Equations \ref{sde3} and \ref{sde4} follow from taking time derivatives of the expressions for $\gamma$ and $\mu$ (see appendix for derivation). Although Equation \ref{sde1} may be replaced by simply updating the energy through the definition of the Lorentz factor in the guiding centre equations (Equation \ref{NRGC3}), this approach was implemented in order to allow generalisation of the scattering model in future work. Our initial choice of the friction and diffusion coefficients $F_\beta$ and $D_{\beta\beta}$ are $F_\beta = -\beta \frac{v_{tot}}{\lambda}$ and $D_{\beta\beta} = (1 - \beta^2)\frac{v_{tot}}{\lambda} $, where the mean free path is parametrised by
\begin{equation}\label{mean-free-path}
\lambda = \lambda_0\left( 1 + \frac{v}{v_{th}}\right)^{\alpha}\kappa,
\end{equation} 
with $\lambda_0 = 2\times 10^8\unit m$, representing the mean free path of an electron in a plasma with coronal parameters. We integrate Equations \ref{sde1} and \ref{sde2} using an Euler scheme whose timestep is the minimum of $dt_0 = 5\times 10^{-9} \unit s$ and $dt_s=1/(3\nu) =  \lambda/(3v_{tot})$. The value for $dt_0$ was determined by comparing results of integrating particle trajectories between the variable timestep code (without scattering) and imposing $F_\beta = D_{\beta\beta} = 0$ with the fixed timestep code. Multiple values of $dt_0$ were evaluated and one was chosen that could accurately reproduce the trajectory given by the variable timestep code. Although a higher order scheme would have been preferable, the dependence of the coefficients on the particle position in the grid would necessitate extra computation of spatial gradients of the fields, which would increase computation time. Furthermore the timestep must be less than the time between scattering events, hence requiring the choice of the minimum of $dt_0$ and $dt_s$. After updating $\beta$ and $\gamma$, we update the magnetic moment and parallel velocity to calculate the position of the guiding centre in Equation~\ref{NRGC1}. The position is then integrated by the Runge-Kutta scheme with the timestep used in the Euler scheme, $dt_0$. 


\section{Results of test particle calculations}
\label{results}

\subsection{Configuration of test particle code}
\label{setup}
\begin{figure}[ht]
        \includegraphics[width = 0.5\textwidth]{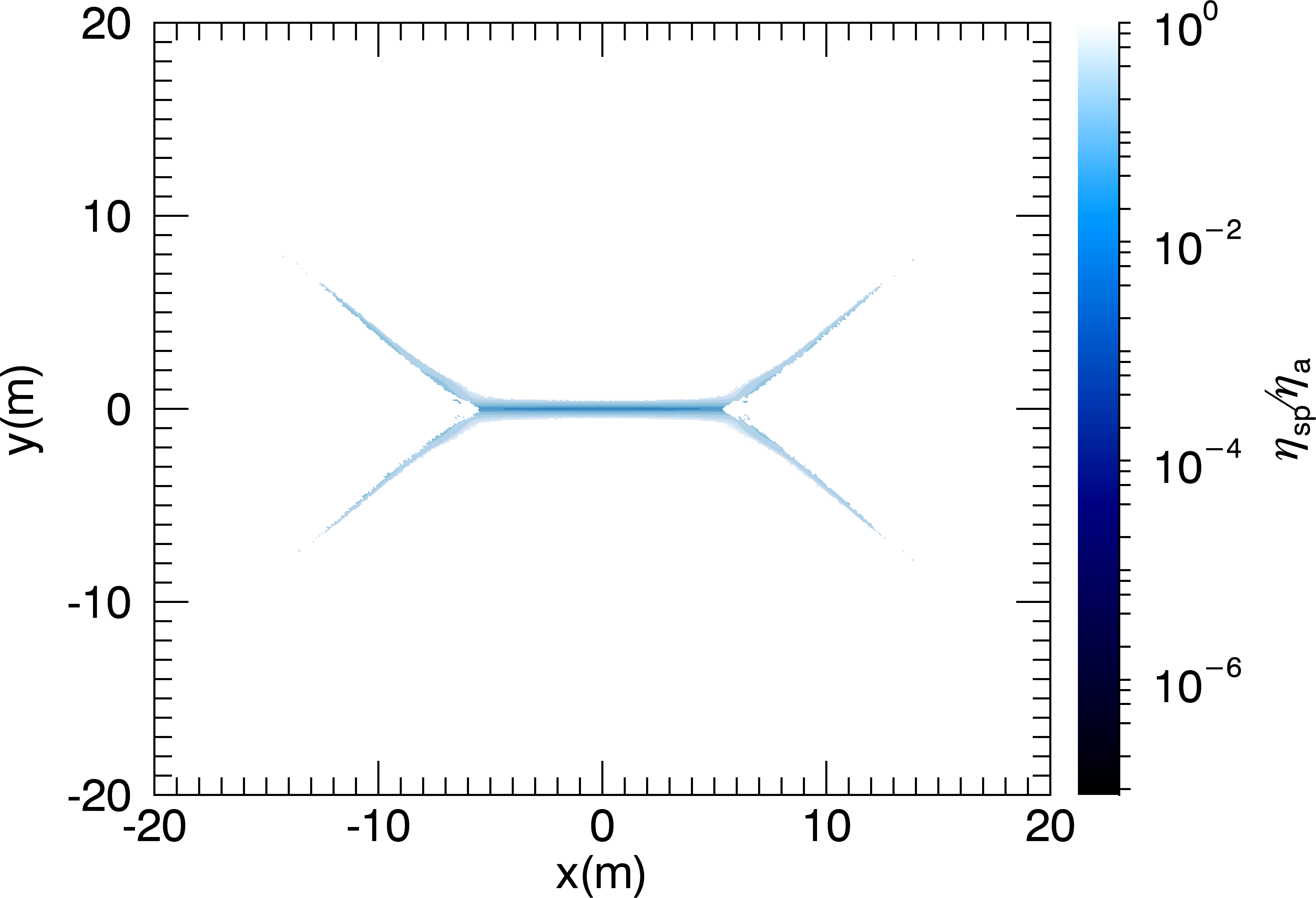}
        \caption{Spatial dependence of the ratio of Spitzer resistivity to anomalous resistivity in the snapshot of the MHD simulation into which test particles are injected. White areas surrounding the current sheet do not have a specified anomalous resistivity, hence the ratio is calculated only where $\eta_a \neq 0$.}
        \label{kappa}
\end{figure}

To study the effect of pitch angle scattering on particle behaviour, we initialise test particle orbits in the MHD snapshot shown in Figure \ref{mhd_snapshot2}, and integrate the governing equations for their evolution, detailed in Section~\ref{particle-eqns}, until the orbit leaves the computational domain. We compare the results of calculations in the presence of different scattering rates by varying the values of $\kappa$ and $\alpha$ in Equation~\ref{mean-free-path}. The parameter $\alpha$ determines how the mean free path changes as a function of test particle velocity, with $\alpha > 0$ resulting in a longer mean free path (and hence less scattering) at higher particle velocities, while $\alpha < 0$ results in a decreasing mean free path for higher velocities. A simple scaling of the mean free path can be applied by varying $\kappa$, with higher values leading to a longer mean free path and less scattering. We introduce a dependence on the anomalous resistivity into the mean free path by setting $\kappa = \eta_{sp}/\eta_a$, where $\eta_{sp}$ is the local Spitzer resistivity at the position of the guiding centre. To get an idea of the spatial dependence of the Spitzer resistivity on position in our MHD simulation, a contour plot of the ratio $\kappa = \eta_{sp}/\eta_a$ is shown in Figure \ref{kappa}.

We perform test particle simulations with the following choices of parameters: to investigate the effect of velocity-dependent scattering we choose $\alpha = \pm 2,0$, with $\kappa = \eta_{sp}/\eta_a$; to investigate the effect of anomalous resistivity we take $\alpha = 0$ and $\kappa = 10^{-5}, 10^{-6}, 2\times 10^{-8},\eta_{sp}/\eta_a$. The mean free path is related to the scattering frequency by $\nu = v_{tot}/\lambda$. In order for the guiding centre approximation to remain valid, the scattering frequency must not exceed the gyrofrequency of the test particle. This restriction on the scattering frequency is dependent on the test particle gyrational velocity, as well as the local magnetic field strength. It is difficult to predict if a test particle orbit will break this condition, however, we find that for values $\kappa < 5\times 10^{-9}$ the scattering frequency starts to regularly exceed the gyrofrequency.
In addition to performing test particle simulations with scattering included at different rates, we perform the same simulations without scattering using the variable timestep 4th order Runge-Kutta code. We refer to these simulations as the control cases. 

To compute test particle energy spectra, we integrate $5\times 10^5$ particle orbits for each of the parameter regimes mentioned above. The particle orbits are distributed with uniformly random initial positions inside a portion of the computation box. This portion is centred on the reconnection region and has a side length of 2 in normalised units (the whole computational box has a side length of 4, also centred on the reconnection region; see Figure \ref{mhd_snapshot2}). The initial pitch angle takes on 100 evenly distributed values between $10^{\circ}$ and $170^{\circ}$ and the initial energy takes on 50 evenly distributed values between $10 \unit{eV}$ and $320\unit{eV}$ (this energy range covers over 90\% of the maxwellian at $10^6\unit K$). These choices mean that there are 100 particle orbits for every combination of initial pitch angle and energy, each having a different (uniformly random) initial position. 

The final energy and position of each orbit is recorded as it exits the computational box. Each orbit is weighted in proportion to the plasma density at its initial position, so that the initial particle energy distribution is approximately a Maxwellian at a temperature of $10^{6}\unit K$ and the initial distribution of the cosine of the pitch angle is uniform. The resulting energy spectra are shown in Figure~\ref{spectra1}.

\subsection{Selected trajectories}
\begin{figure*}[ht]
\begin{subfigure}[b]{0.32\textwidth}
\includegraphics[width = \textwidth]{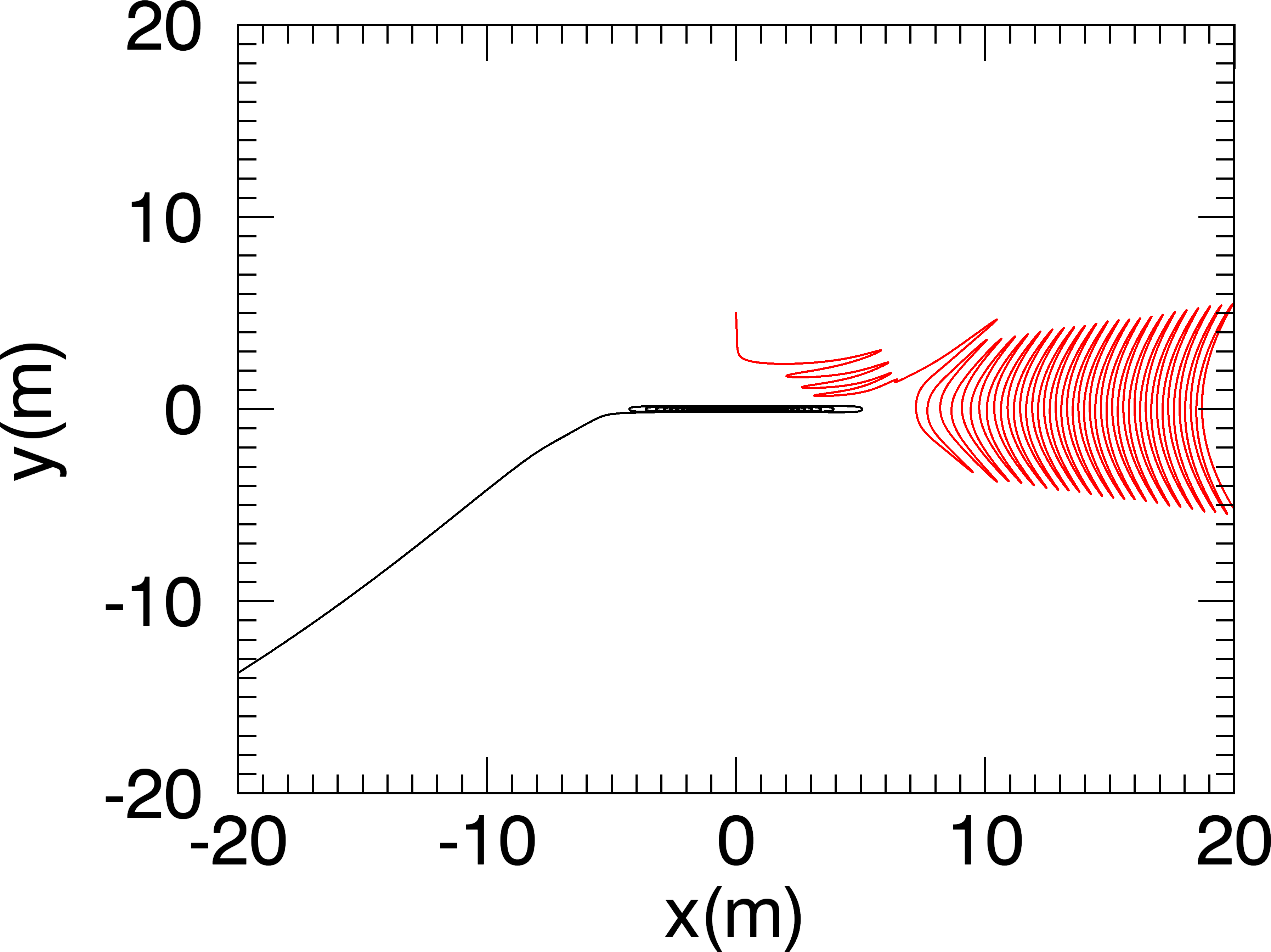}
\caption{}\label{orbits1a}
\end{subfigure}
\begin{subfigure}[b]{0.32\textwidth}
\includegraphics[width = \textwidth]{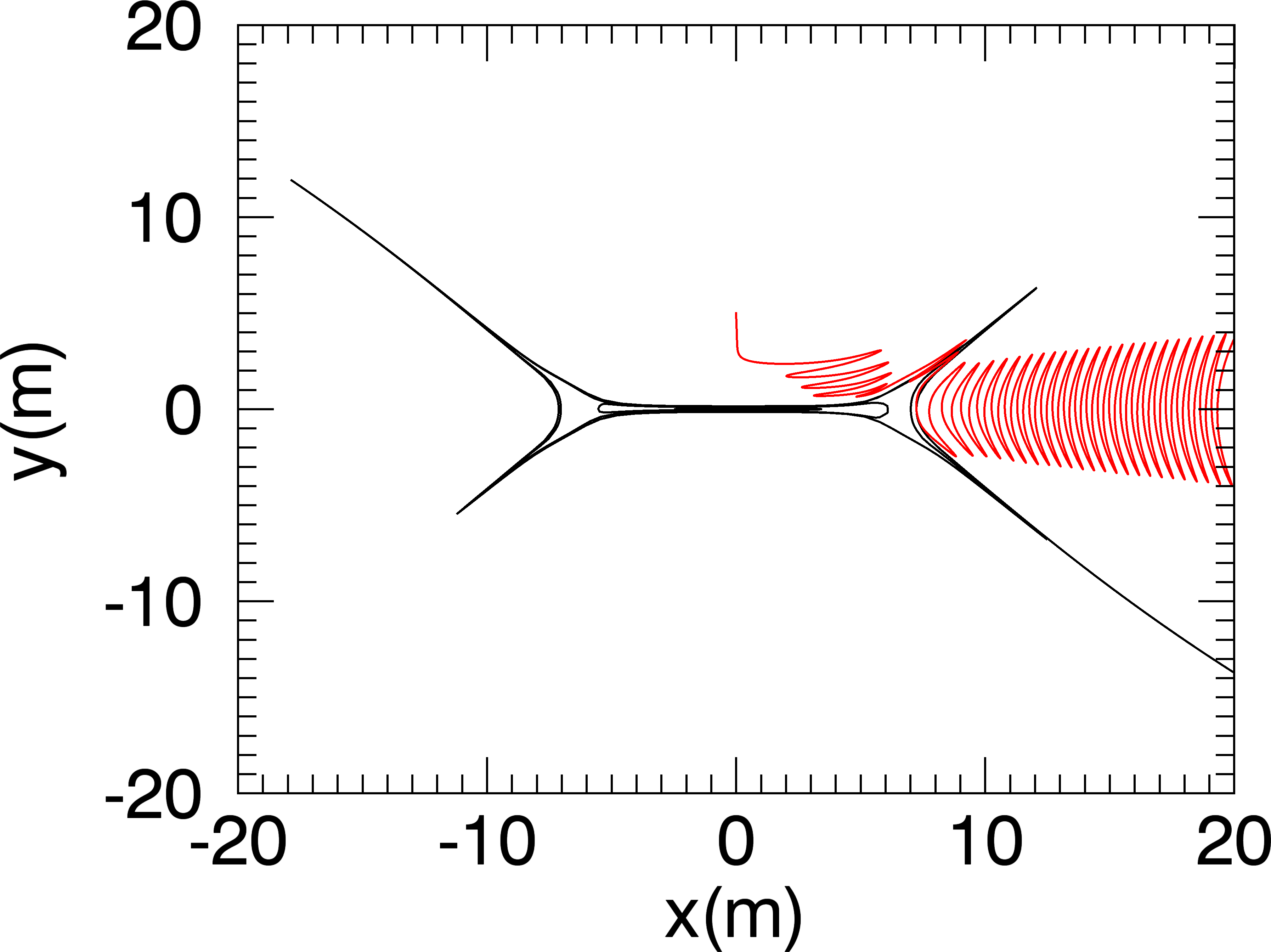}
\caption{}\label{orbits1b}
\end{subfigure}
\begin{subfigure}[b]{0.32\textwidth}
\includegraphics[width = \textwidth]{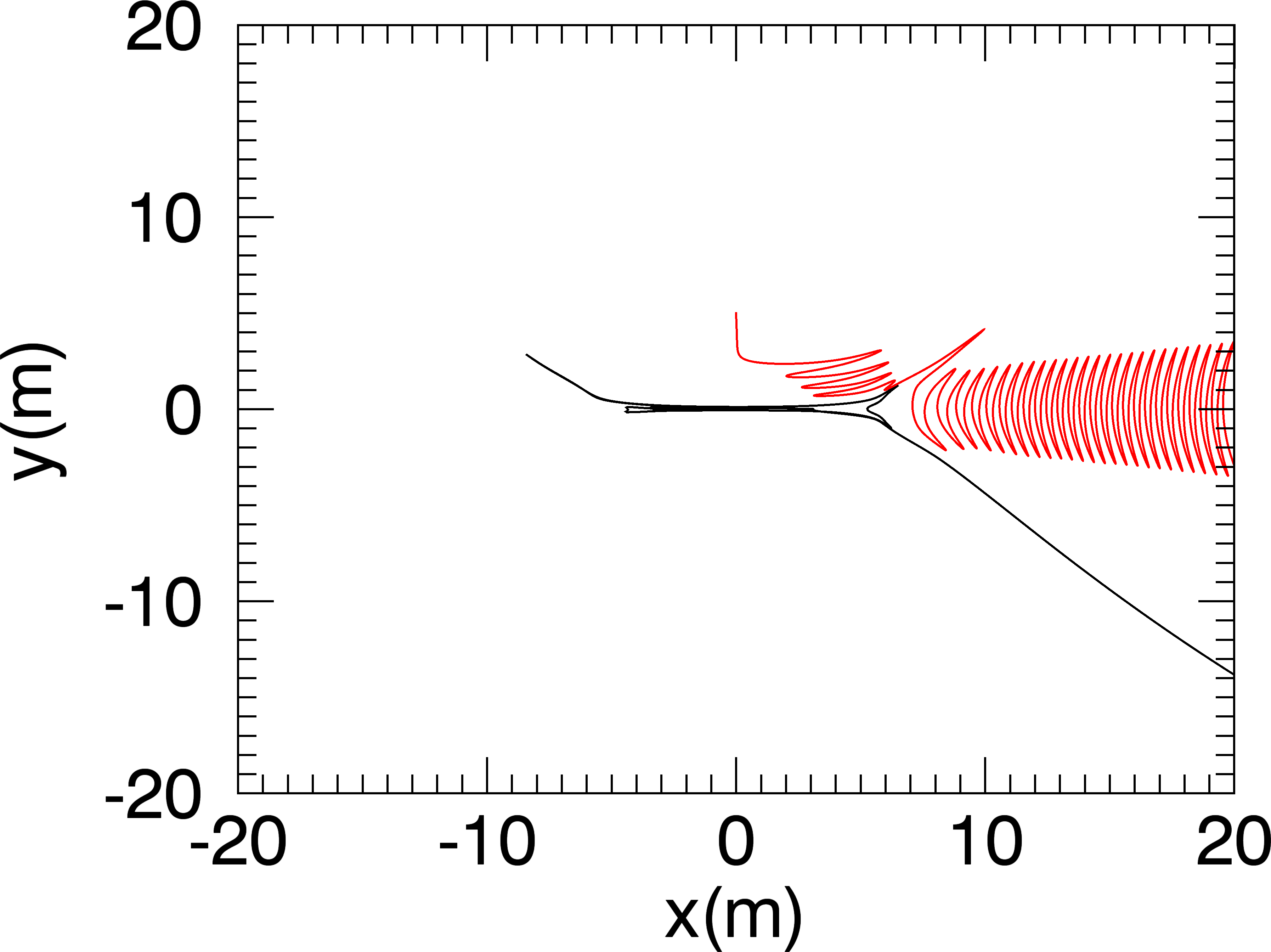}
\caption{}\label{orbits1c}
\end{subfigure}
\caption{Orbit trajectories for test particles initialised at $(x,y) = (0,0), (0,5) \unit m$ (black and red trajectories respectively) within the MHD snapshot. The initial pitch angle is $90^\circ$ and kinetic energy $320\unit{eV}$. Test particle orbit calculations were performed \subref{orbits1a} without scattering, \subref{orbits1b} with scattering where $\kappa = 10^{-6}$, and \subref{orbits1c} where $\kappa = \eta_{sp}/\eta_a$.}\label{orbits1}
\end{figure*}

\begin{figure*}[ht]
\begin{subfigure}[b]{0.49\textwidth}
\includegraphics[width = \textwidth]{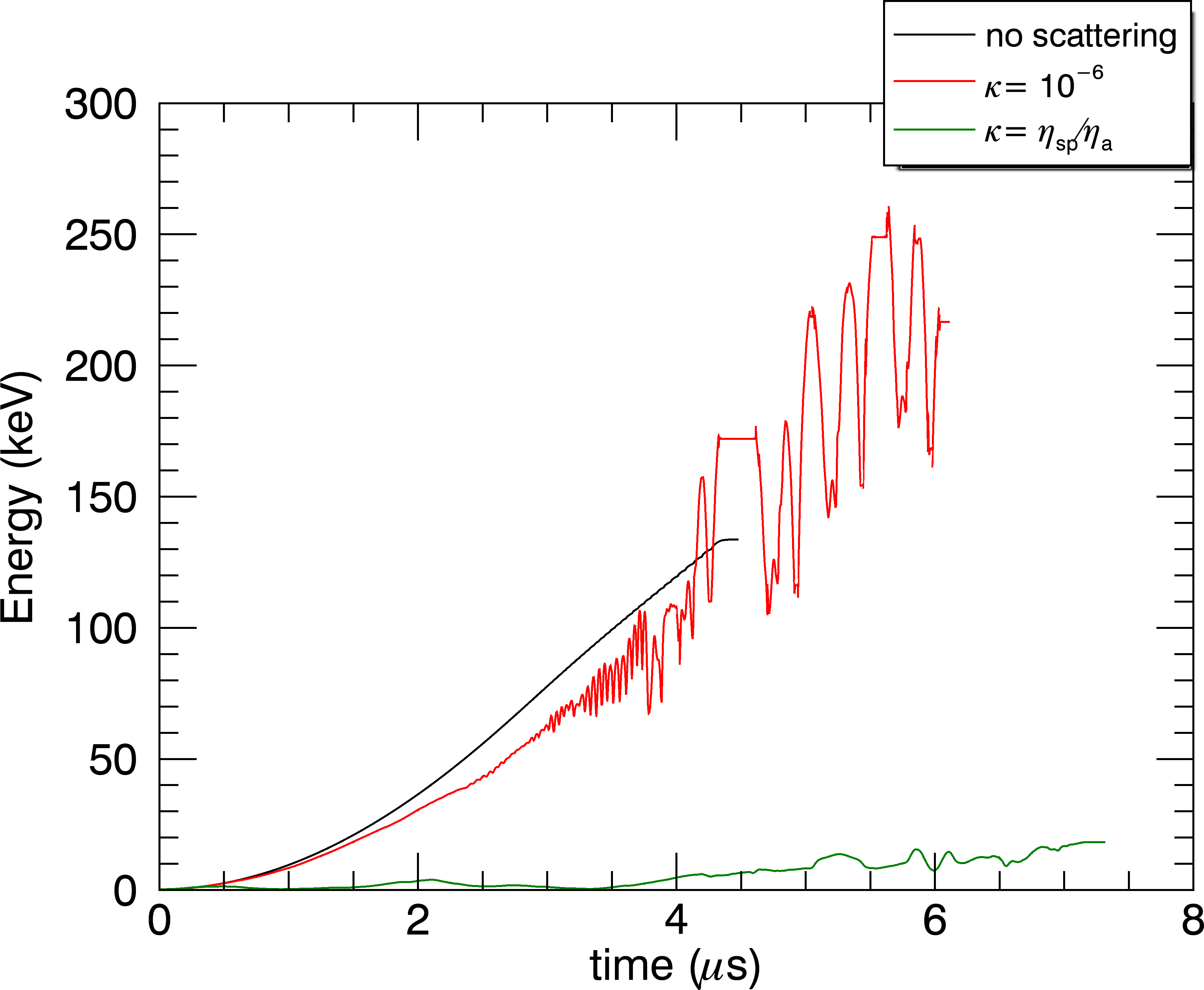}
\caption{}\label{orbits2a}
\end{subfigure}
\begin{subfigure}[b]{0.49\textwidth}
\includegraphics[width = \textwidth]{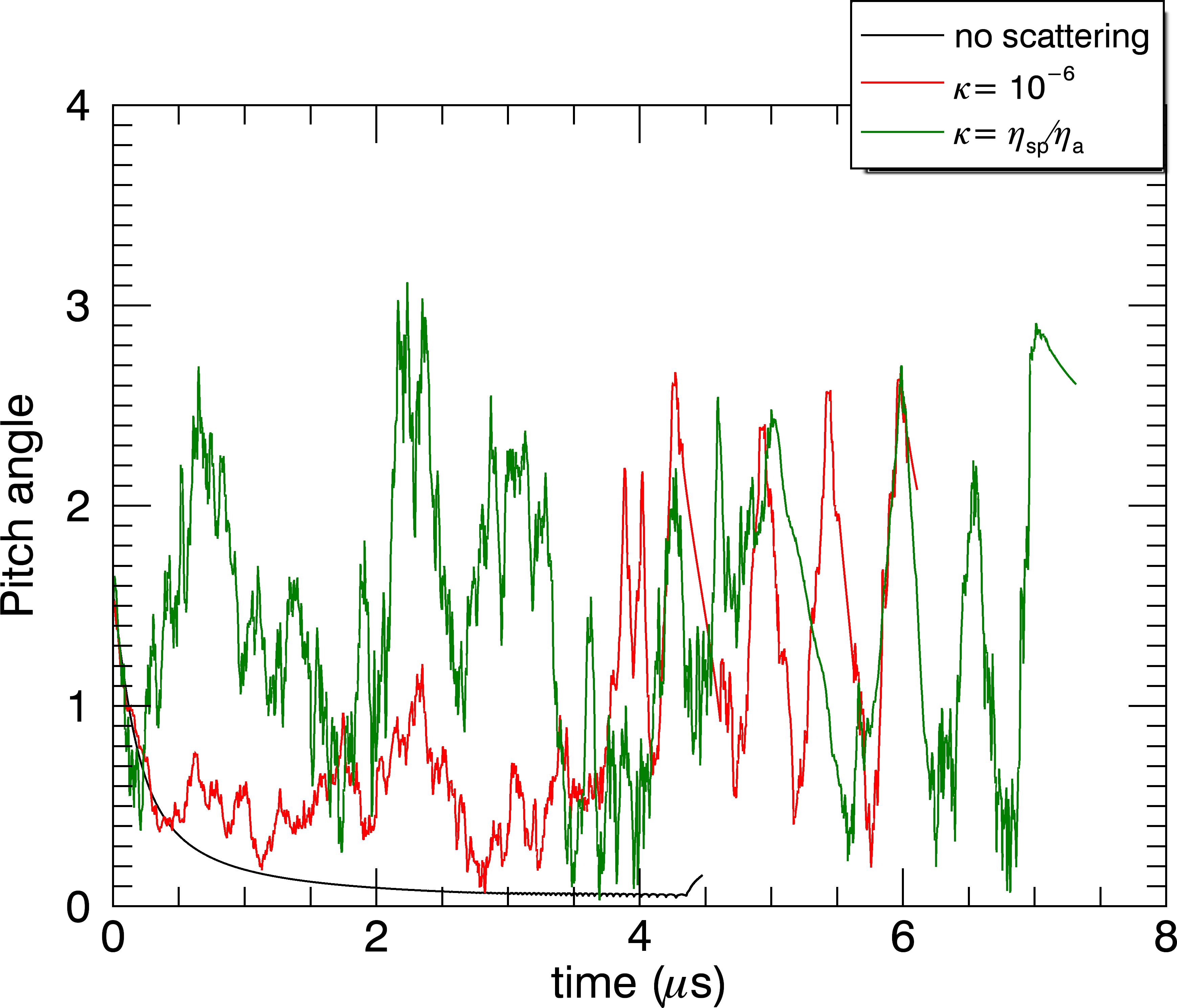}
\caption{}\label{orbits2b}
\end{subfigure}
\\
\begin{subfigure}[b]{0.49\textwidth}
\includegraphics[width = \textwidth]{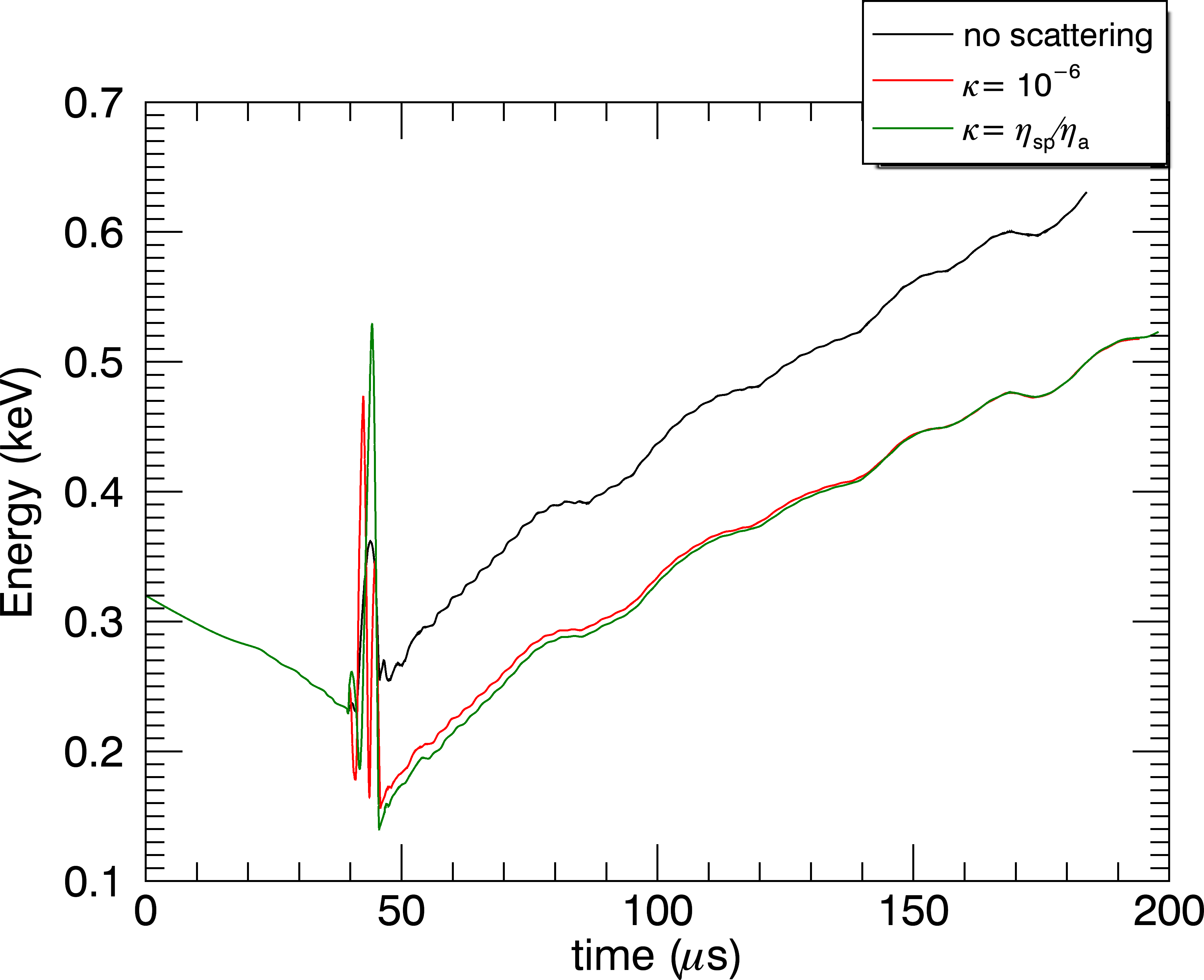}
\caption{}\label{orbits2c}
\end{subfigure}
\begin{subfigure}[b]{0.49\textwidth}
\includegraphics[width = \textwidth]{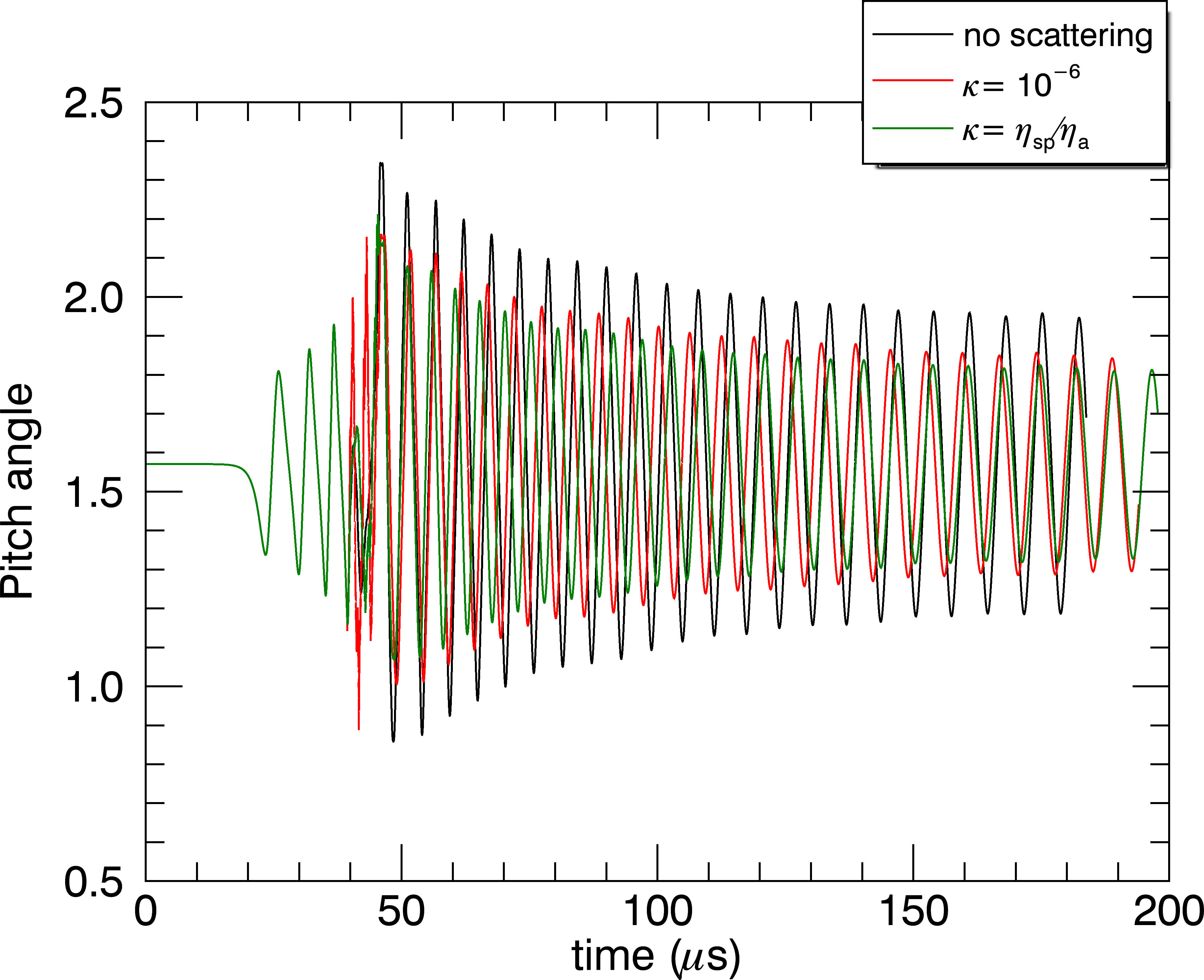}
\caption{}\label{orbits2d}
\end{subfigure}
\caption{Orbit energy and pitch angle evolution for the trajectories calculated above. Panels \subref{orbits2a} and \subref{orbits2b} refer to orbits initialised within the current sheet (i.e. for $y = 0\unit m$), while panels \subref{orbits2c} and \subref{orbits2d} refer to orbits initialised outside of the current sheet (at $y = 5\unit m$).}\label{orbits2}
\end{figure*}

Our primary interests are the energy spectra obtained through many orbit calculations, however, it is initially enlightening to examine selected orbit trajectories, energy, and pitch angle evolution. Such examples reveal the general effect of pitch angle scattering on individual orbits. To do this we place test particles at two distinct initial positions, $y_0 = 0$ and $5 \unit m$ (in both cases with $x = 0 \unit m$), with initial pitch angle $\theta_0 = 90^{\circ}$ and kinetic energy is $320 \unit{eV,}$ into the MHD snapshot. These initial conditions are chosen so that the effect of scattering is evident on orbits that drifts into the reconnection region due to the $\textbf E \times \textbf B$ drift, as well as for orbits starting within the reconnection region. The particle trajectories are calculated as described in the previous section with no scattering, scattering with $\kappa = 10^{-6}$ in Equation \ref{mean-free-path}, and with $\kappa = \eta_{sp}/\eta_a$. The resulting trajectories, energy evolution, and pitch angle evolution are shown in Figures \ref{orbits1} and \ref{orbits2}. 

Due to the magnetic moment no longer being conserved in the case of the different scattering regimes, the orbit trajectories in Figures \ref{orbits1b} and \ref{orbits1c} differ from the control case (Figure \ref{orbits1a}). This is caused by terms in the guiding centre equations (Equations \ref{NRGC1}, \ref{NRGC2}) proportional to $\mu$ having a randomising effect on the particle drifts when scattering is included (as $\mu$ is no longer constant). For the particle orbit initialised in the current sheet, the more chaotic evolution of the pitch angle when $\kappa = \eta_{sp}/\eta_a$ (see green curve in Figure \ref{orbits2b} in comparison to the red and black curves) suggests that this choice of $\kappa$ produces stronger scattering within the diffusion region than if $\kappa = 10^{-6}$. When $\kappa = 10^{-6}$, we note that the particle orbit crosses the reconnection region multiple times (see black particle orbits in Figures \ref{orbits1b} and \ref{orbits1c}), as has been reported previously \citep[see][]{burge-et-al2014}, which is an effect that cannot happen in the absence of scattering. Orbits which enter the current sheet multiple times can traverse a greater potential drop than if they were evolving deterministically, and hence gain more energy. Due to the stochastic nature of the orbit, such behaviour and associated increased energy is not guaranteed even with identical orbit initial conditions. We anticipate that the presence of scattering will yield energy spectra containing higher maximum energies than the case without scattering, as a result of particle trajectories traversing the reconnection region multiple times. 

Orbits which start outside of the reconnection region are not subject to as much scattering and acceleration if they drift into the separatrices rather than the central current sheet. As a result, although some scattering is evident in the trajectory (red lines in Figure \ref{orbits1}) and pitch angle evolution (Figure \ref{orbits2d}) of the particle orbits initialised at $y = 5\unit m$, energy changes at the end of the orbit are much less evident than for the particle orbits initialised inside the current sheet.

\subsection{Energy spectra}
\label{spectra}

\begin{figure*}[ht]
\begin{subfigure}[b]{0.49\textwidth}
\includegraphics[width = \textwidth]{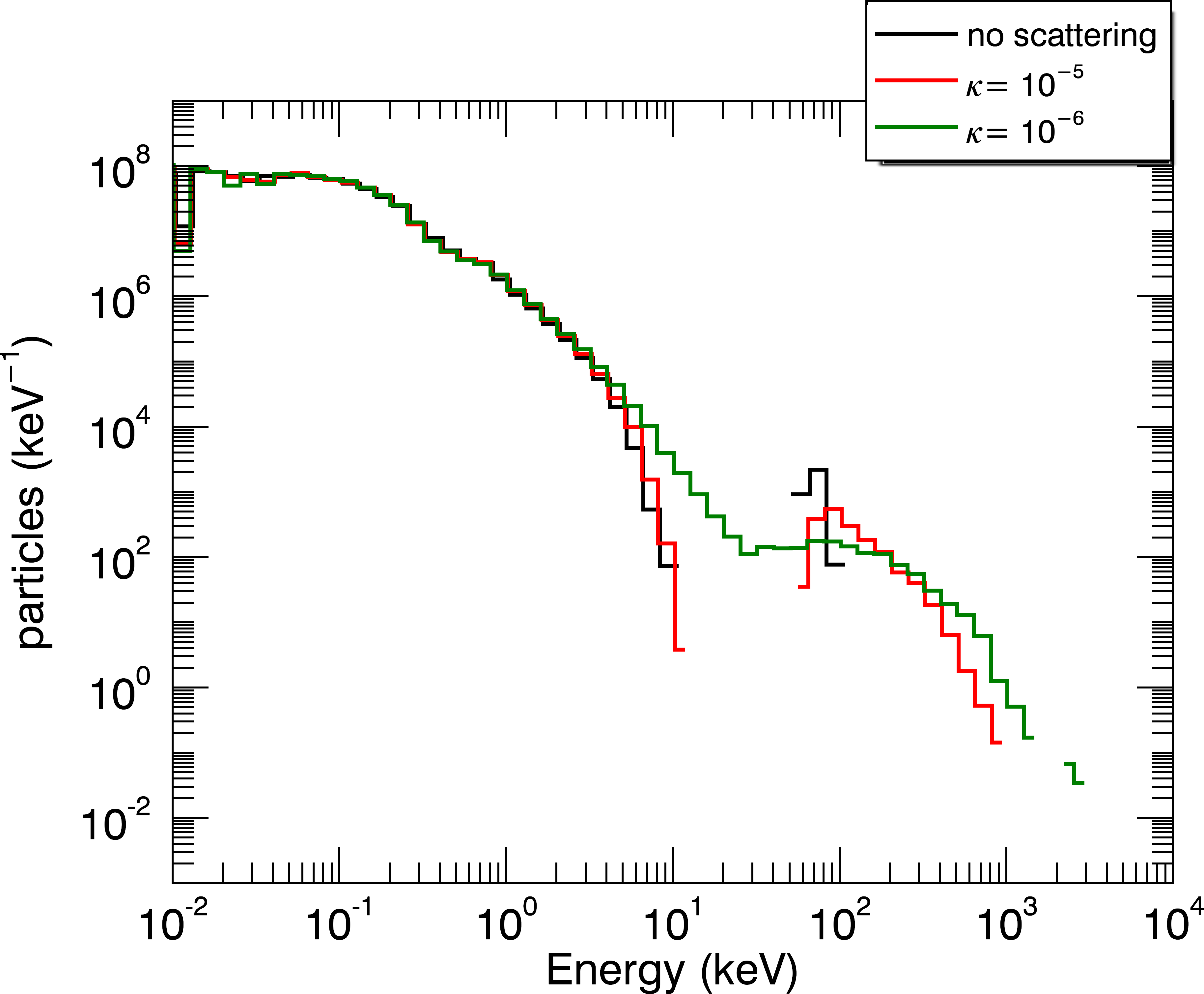}
\caption{}\label{spectra1a}
\end{subfigure}
\begin{subfigure}[b]{0.49\textwidth}
\includegraphics[width = \textwidth]{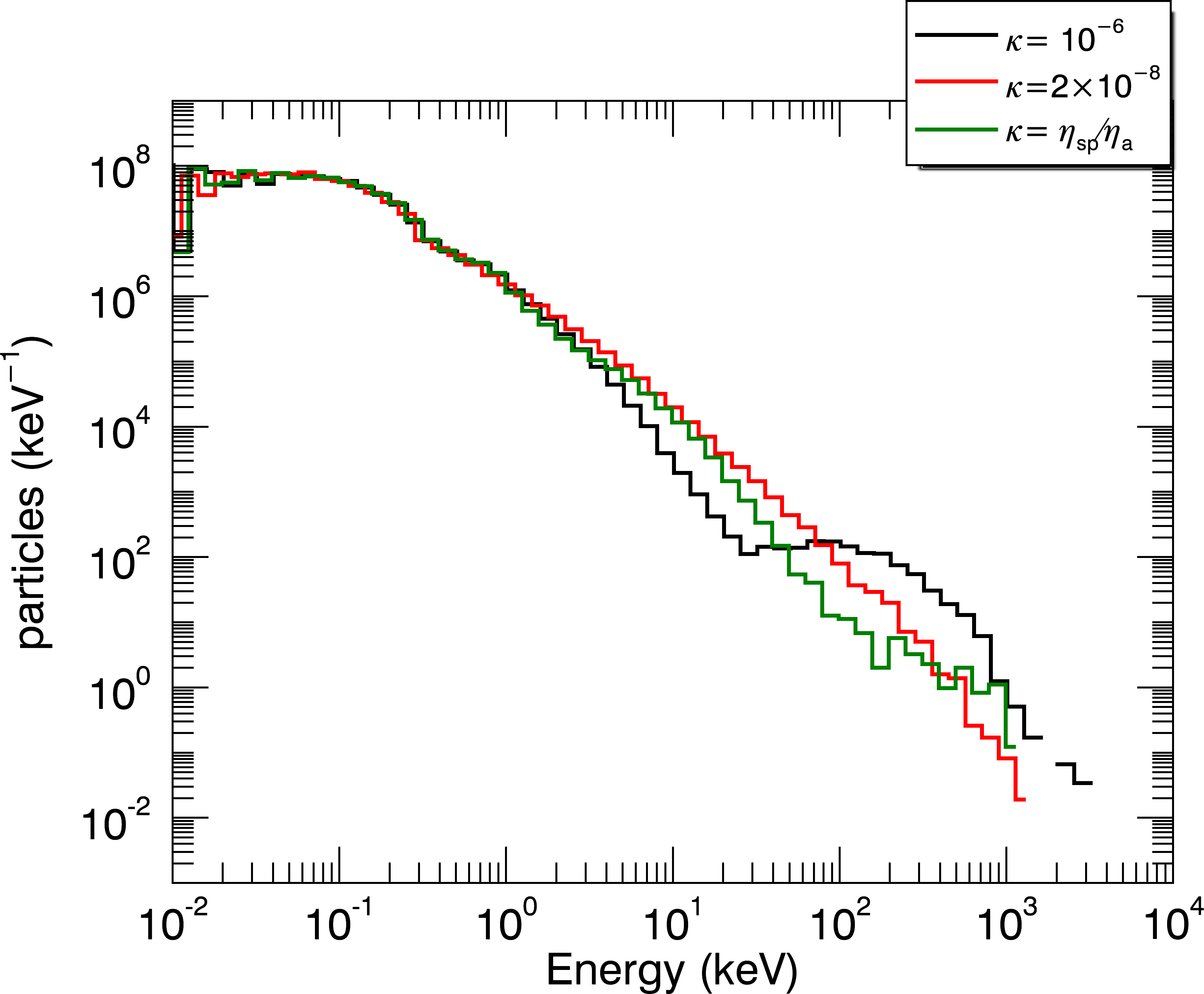}
\caption{}\label{spectra1b}
\end{subfigure}
\\
\begin{subfigure}[b]{0.49\textwidth}
\includegraphics[width = \textwidth]{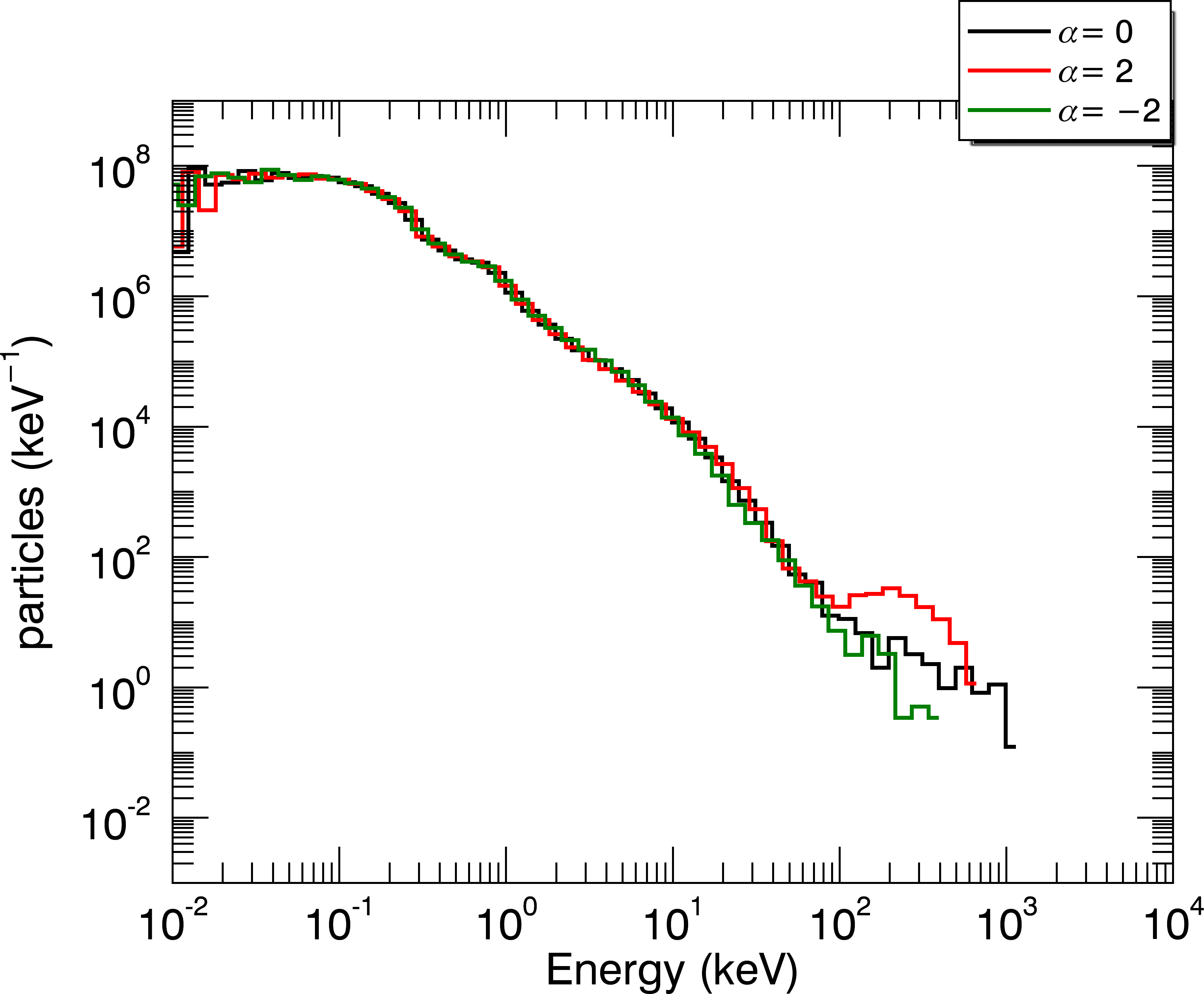}
\caption{}\label{spectra1c}
\end{subfigure}
\begin{subfigure}[b]{0.49\textwidth}
\includegraphics[width = \textwidth]{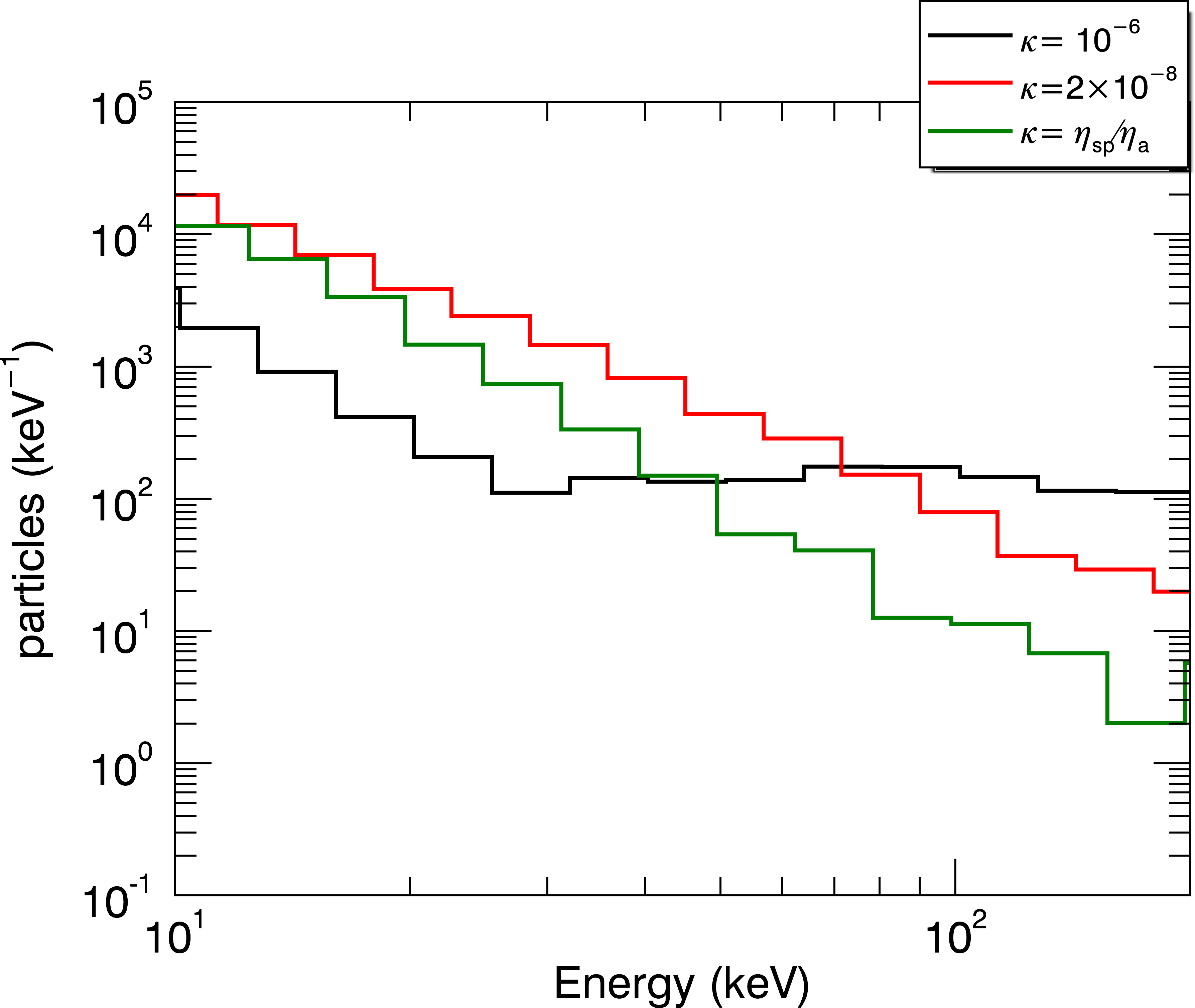}
\caption{}\label{spectra1d}
\end{subfigure}
\caption{Final test particle energy spectra for various scattering models. In all cases $5\times 10^5$ test particle orbits are calculated, then each orbit is weighted in proportion to the local density at the initial position of the orbit to ensure the initial energy distribution is a Maxwellian at $T = 10^6\unit K$, and that the initial pitch angle cosine distribution is uniform. Subsection \ref{setup} shows the initial conditions of the simulations. In panels a and b we set $\alpha = 0$ and vary the value of $\kappa$, whereas in panel c we take $\kappa = \eta_{sp}/\eta_a$ and vary $\alpha$. Panel d shows the spectra from panel b between 1 and 100 \unit{keV}.}
\label{spectra1}
\end{figure*}

In Figure \ref{spectra1a} we compare the spectra produced by the control case (without scattering, black curve), with the scattering cases where $\kappa = 10^{-5}, 10^{-6}$ (in both of these we set $\alpha = 0$). We note that there is a break in the spectrum of the control case. A small population of highly accelerated particles achieve energies of approximately 100 \unit{keV} (approximately 0.3\% of the total number of orbits, after weighting). This break in the spectrum is due to the small size of the reconnection region. When scattering is introduced, with $\kappa = 10^{-5}$ (red curve in Figure \ref{spectra1a}), there is an increase in the spread of energies obtained by the highly energised particles orbits (compared to the control case), while the general shape of the spectrum remains unchanged. The spectrum of the scattering case with $\kappa = 10^{-6}$ (green curve in Figure \ref{spectra1a}) is smoother, without any breaks in the spectrum. This suggests that scattering is much more effective for smaller values of $\kappa$. Both green and red curves in Figure \ref{spectra1a} contain significant numbers of particle orbits achieving energies much greater than the maximum energy achieved by any particle orbit in the control case (in both scattering regimes approximately 0.15\% of particle orbits achieve energies higher than any unscattered orbit, corresponding to approximately half of the total highly accelerated population in the control case).

Next we compare spectra produced with $\kappa$ given by $\kappa = \eta_{sp}/\eta_{a}$ to the constant $\kappa = 10^{-6}$ and $\kappa = 2\times 10^{-8}$ cases, again with $\alpha = 0$ (see Figure~\ref{spectra1b}). Figure~\ref{spectra1d} shows a restricted energy range (between 1 and 200 \unit{keV}) of the same spectra. The value of $\kappa = 2\times 10^{-8}$ is chosen to be comparable to the minimum value of $\eta_{sp}/\eta_a$ (since $\eta_{sp} \propto T^{-3/2}$ this is the location in the MHD simulation with the highest temperature, i.e. in the middle of the current sheet). Since all three cases examined here include relatively strong scattering, we see that there are no breaks in any spectrum, and furthermore there are more particles with energies in the region of 10 \unit{keV} in the case when $\kappa = \eta_{sp}/\eta_{a}$ and $\kappa = 2\times 10^{-8}$ than when $\kappa = 10^{-6}$, with fewer higher energy particles (in particular 1.6\% of the total particle orbits have energies between 5 and 30 \unit{keV} for the case $\kappa = \eta_{sp}/\eta_a$, compared to 2.3\% for the $\kappa = 2\times 10^{-8}$ case and 0.5\% for the $\kappa = 10^{-6}$ case). The dependency of the mean free path on the resistivity leads to lower maximum energies than even a constant but lower value of $\kappa$. This is because the ratio $\eta_{sp}/\eta_a$ increases drastically in the separatrices where the temperature is lower, resulting in fewer particles being scattered. The absence of scattering within the separatrices means that fewer orbits are able to repeatedly cross the acceleration region, resulting in lower energies. We also note that scattering with $\kappa = 2\times 10^{-8}$ yields fewer particles at energies above 100\unit{keV} when compared with the $\kappa = 10^{-6}$ case. In both cases, the maximum energy obtained by particles is still higher than for the case without scattering. 

Finally, in Figure~\ref{spectra1}c, we consider spectra produced by varying the velocity dependence of the scattering model. In Figures \ref{spectra1a} and \ref{spectra1b} we fixed $\alpha = 0$ and varied values of $\kappa$. Now we set $\kappa = \eta_{sp}/\eta_{a}$ and consider $\alpha = -2,0,2$. There is a very small difference in the spectra above 1\unit{keV,} with the $\alpha = -2$ case having very slightly more particle orbits at lower energies (8.1\% of total particle orbits with energies between 1 and 10 \unit{keV}, as opposed to 7.9\% and 7.6\% for the $\alpha = 0$ and $\alpha = 2$ cases respectively) and fewer higher energy orbits (0.004\% of total particle orbits with energies greater than 100 \unit{keV}, as opposed to 0.01\% and 0.04\% for the $\alpha = 0$ and $\alpha = 2$ cases respectively). The spectrum for the $\alpha = 0$ case falls in between the other two. This indicates stronger scattering occurring for large negative $\alpha$. This is to be expected as the mean free path decreases for large ratios of the particle velocity to the thermal velocity, implying more scattering. The small difference between three values of $\alpha$ is due to the factor $1 + v_{tot}/v_{th}$ only varying between approximately 1 and 6 for a test particle starting at the centre of the dissipation region (where the temperature and electric field are at their maximum). Changing the mean free path by several orders of magnitude when varying $\kappa$ has a much greater impact on the spectrum than a change in the velocity dependence.

\begin{figure*}[ht]
\begin{subfigure}[b]{0.49\textwidth}
\includegraphics[width = \textwidth]{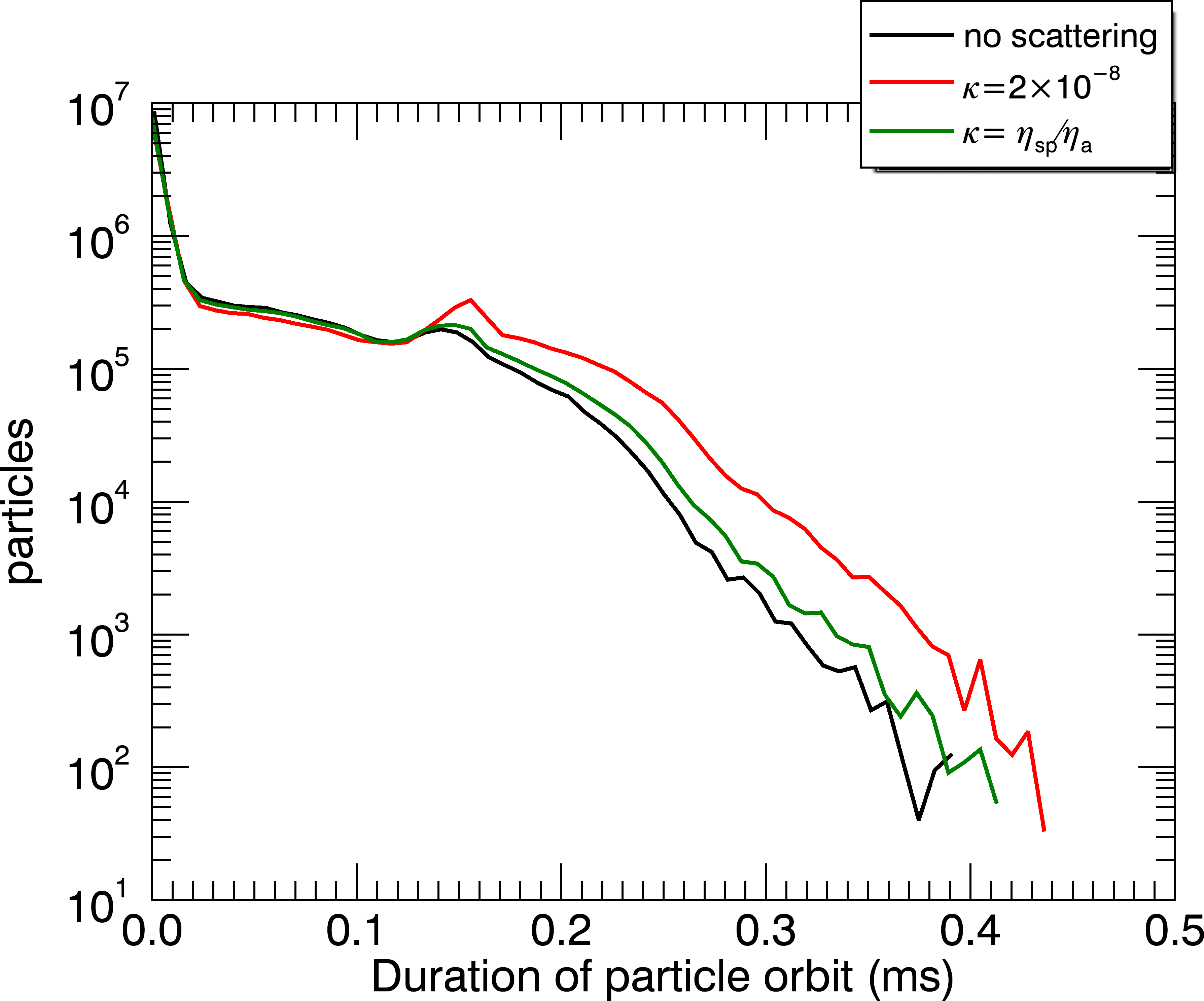}
\caption{Histogram of particle orbit duration}\label{time1a}
\end{subfigure}
\begin{subfigure}[b]{0.49\textwidth}
\includegraphics[width = \textwidth]{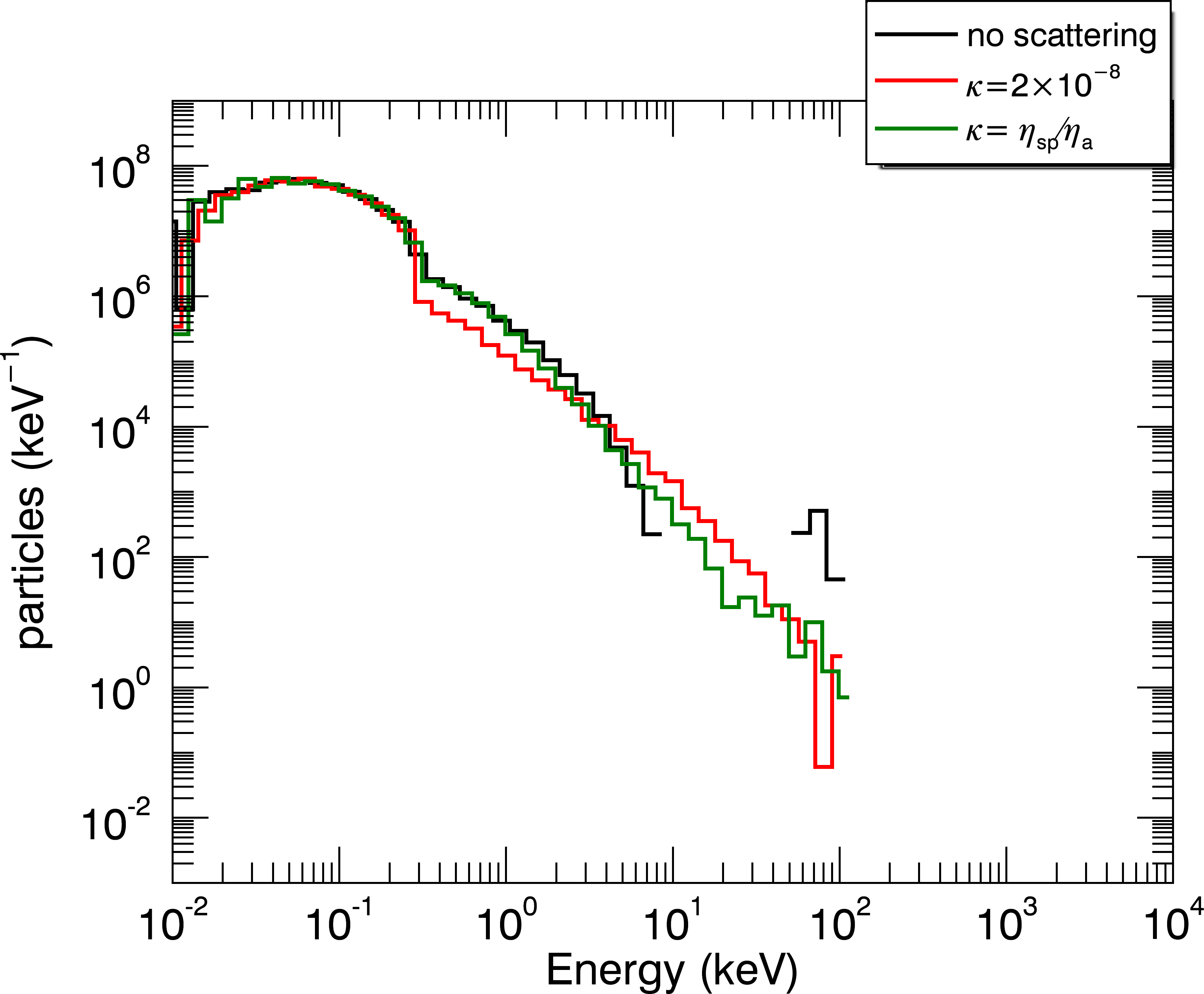}
\caption{$t < 10^{-5}\unit s$}\label{time1b}
\end{subfigure}
\\
\begin{subfigure}[b]{0.49\textwidth}
\includegraphics[width = \textwidth]{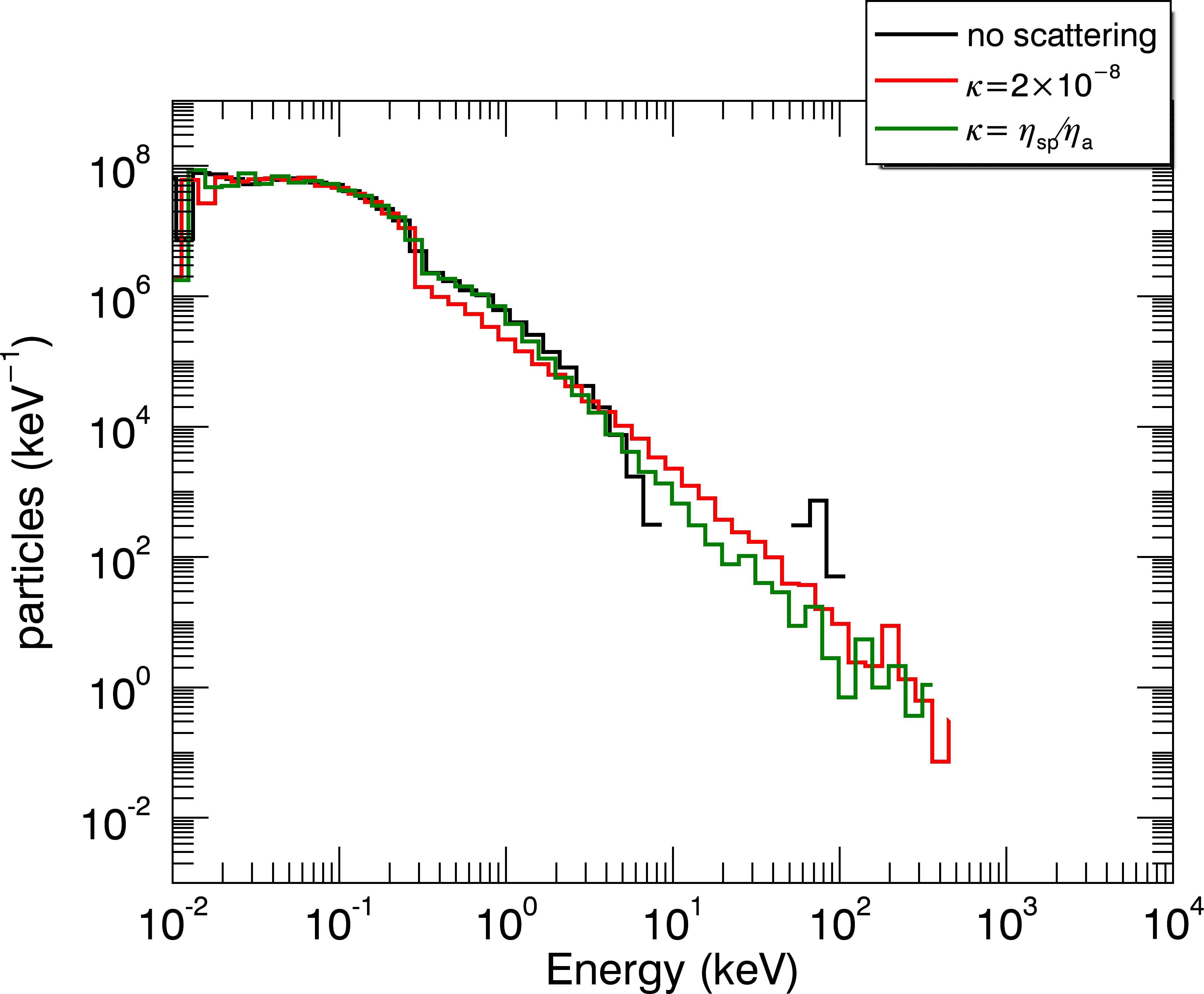}
\caption{$t < 2\times 10^{-5}\unit s$}\label{time1c}
\end{subfigure}
\begin{subfigure}[b]{0.49\textwidth}
\includegraphics[width = \textwidth]{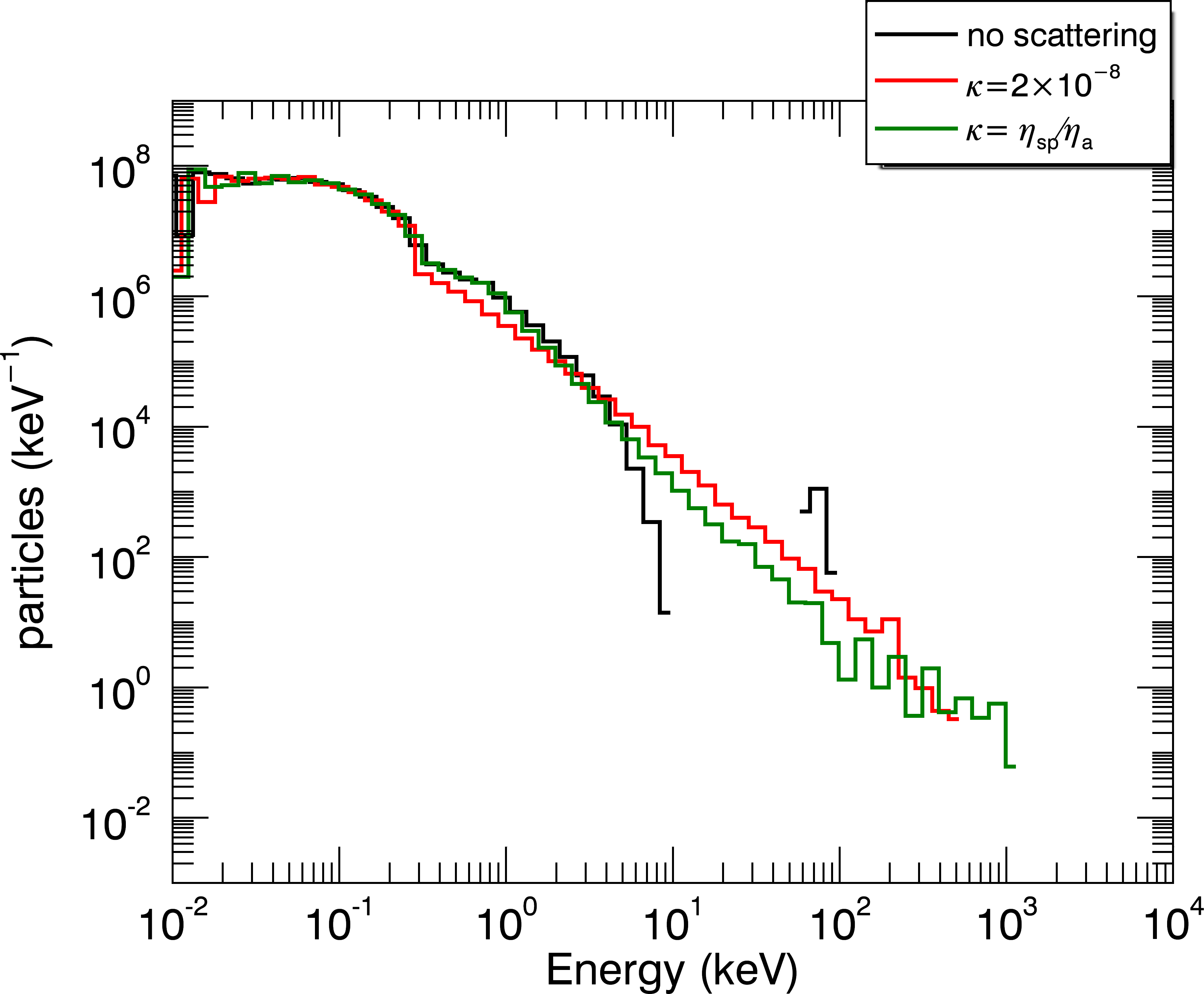}
\caption{$t < 4\times 10^{-5}\unit s$}\label{time1d}
\end{subfigure}
\\
\begin{subfigure}[b]{0.49\textwidth}
\includegraphics[width = \textwidth]{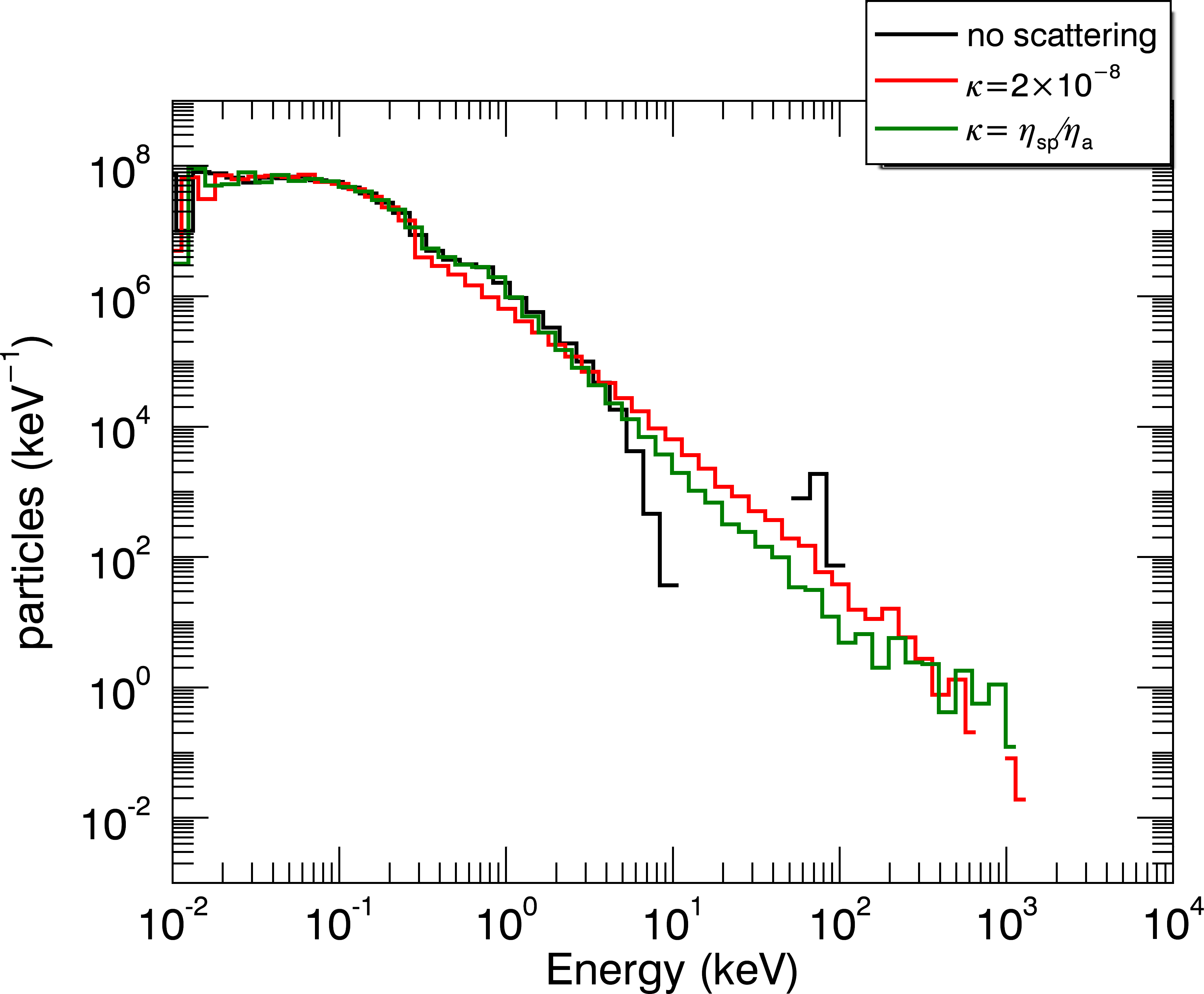}
\caption{$t < 10^{-4}\unit s$}\label{time1e}
\end{subfigure}
\begin{subfigure}[b]{0.49\textwidth}
\includegraphics[width = \textwidth]{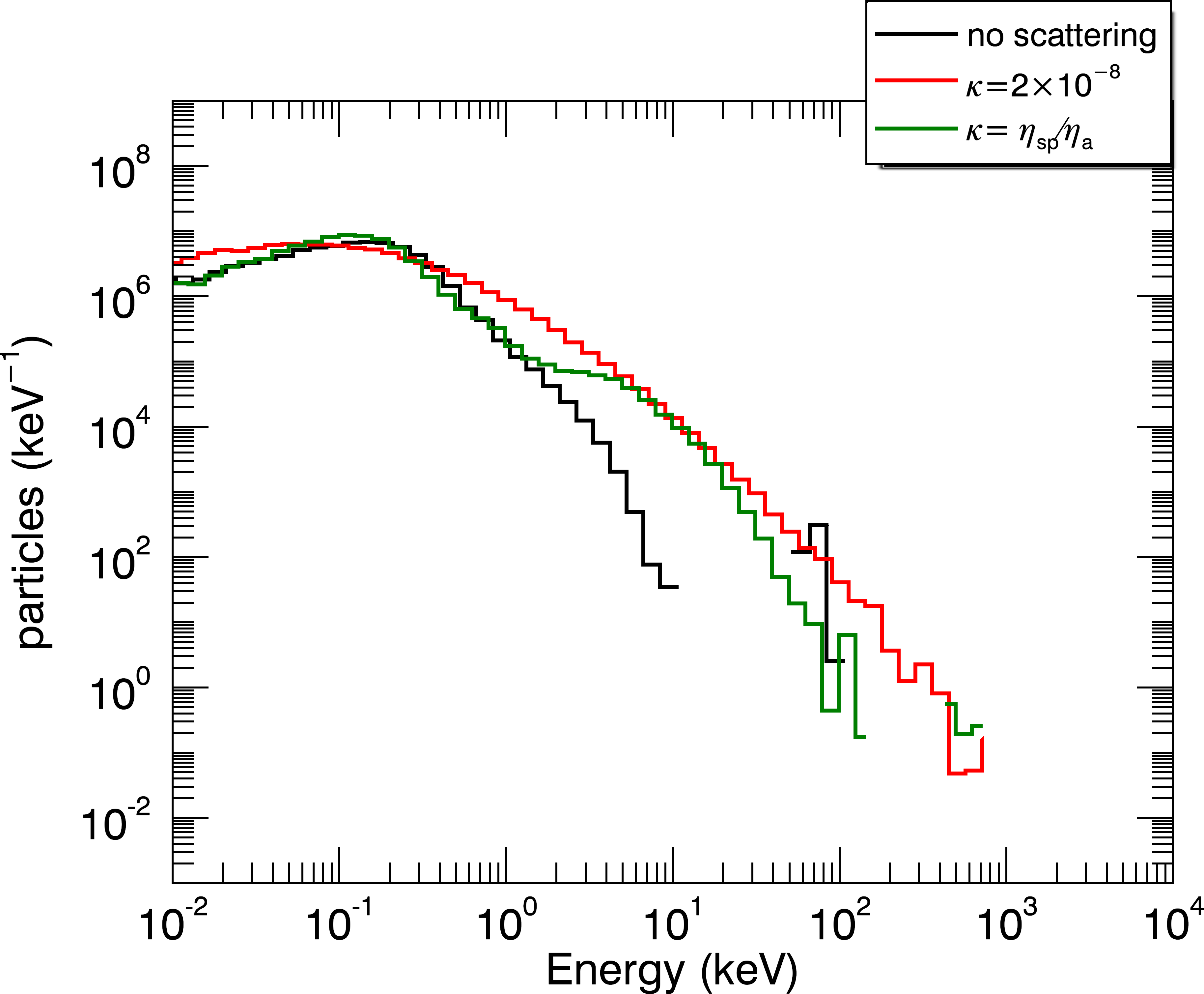}
\caption{$t > 10^{-4}\unit s$}\label{time1f}
\end{subfigure}
\caption{Panel (a) shows a histogram of the duration of the particle orbits for the simulations without scattering, and with scattering where $\kappa = 2\times 10^{-8}$ and $\kappa = \eta_{sp}/\eta_a$. Panels (b-f) show spectra consisting of particle orbits with durations for the indicated time range.}\label{time1}
\end{figure*}

The presence of pitch angle scattering should decrease the rate at which particles are accelerated. In Figure \ref{time1a} we plot a histogram of orbit durations in cases without scattering (black curve), with scattering where $\kappa = 2\times 10^{-8}$ (red curve) and $\kappa = \eta_{sp}/\eta_a$ (green curve, in both of the scattering cases $\alpha = 0$). We see that the number of particles per duration only varies between the three cases above 0.1\unit{ms} durations. The number of particle orbits with duration greater than $0.1\unit{ms}$ is about 14\% for the no scattering case, rising to 16\% for the scattering case where $\kappa = \eta_{sp}/\eta_a$ and 22\% when $\kappa = 2\times 10^{-8}$. Figures \ref{time1b} - \ref{time1f} show orbit spectra with successively longer durations. We note that the spectra of the simulations including scattering extend to progressively higher energies when particle orbits with progressively longer durations are considered. This is again due to particle orbits needing multiple traverses of the current sheet in order to gain energies higher than those possible in the absence of scattering. The abrupt step in the spectra in Figures \ref{time1b}-\ref{time1e} at approximately 320 \unit{keV} is due to the particle orbits which exit the computational box without having encountered the reconnection region. This happens relatively quickly (the exact orbit duration would depend on the initial pitch angle, position, and kinetic energy of each particle orbit, but in all cases occurs faster than 0.1\unit{ms}) and, as such, these particle orbits are not present in Figure \ref{time1f}, resulting in a much smoother spectrum. It is interesting to note the presence of a distinct shoulder starting at energies of approxmately 20\unit{keV} in Figure \ref{time1f} for the $\kappa = \eta_{sp}/\eta_a$ spectrum. Given the already small number of particle orbits which last longer than 0.1\unit{ms}, it is not surprising that this feature is not seen in the full spectrum in Figure \ref{spectra1b}. 

\subsection{Particle orbit escape positions}
\begin{figure}[ht]
\begin{subfigure}[b]{0.49\textwidth}
\includegraphics[width = \textwidth]{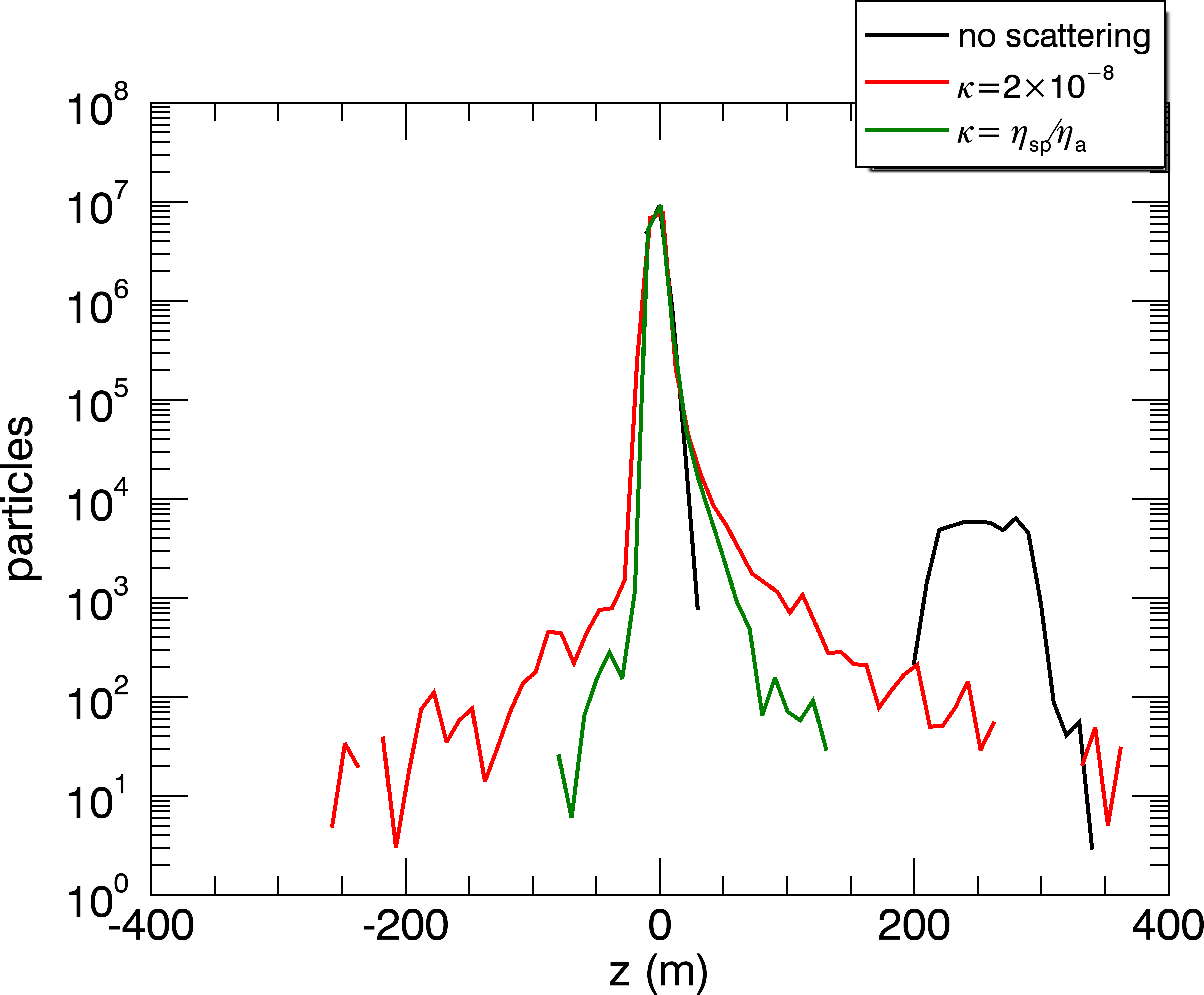}
\caption{}\label{positions1a}
\end{subfigure}
\begin{subfigure}[b]{0.49\textwidth}
\includegraphics[width = \textwidth]{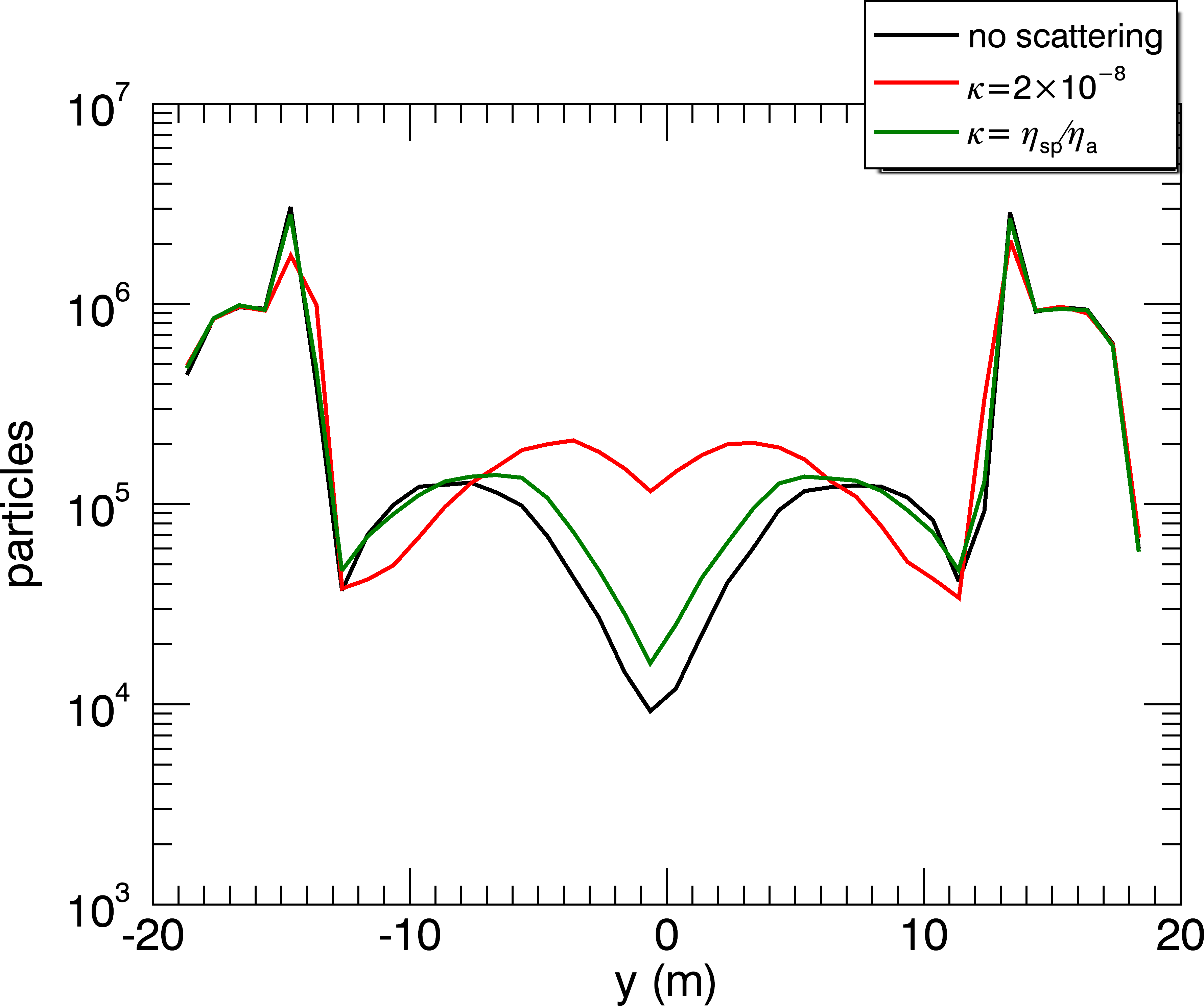}
\caption{}\label{positions1b}
\end{subfigure}
\caption{Histograms of the $y$-, and $z$-positions of particle orbit escape from computational box in the absence of scattering and for the $\kappa = 2\times 10^{-8}$ and $\kappa = \eta_{sp}/\eta_a$ scattering regimes. In both scattering cases, $\alpha$ is set to zero.}
\label{positions1}
\end{figure}

We now turn our attention to the impact of scattering on the final positions of each test particle orbit upon exiting the computational box. In Figure \ref{positions1} we produce histograms for the final $z$ and $y$ positions. We do this for the scattering model when $\kappa = \eta_{sp}/\eta_{a}$ (green curve) and $\kappa = 2\times 10^{-8}$ (red curve), in both cases with $\alpha = 0$, in addition to the control case (black curve). In the control case the highly accelerated particle population primarily escapes the simulation domain between $z = 200$ and $z = 300 \unit m$ causing a prominent increase seen on the right hand side of Figure \ref{positions1a}. In contrast, scattering results in more spread in the final $z$-position. The two scattering models differ in the distribution of the particle orbits final $z$-position, with scattering in the $\kappa = \eta_{sp}/\eta_a$ model resulting in a narrower range of exit locations, compared to the stronger scattering case, with $\kappa = 2\times 10^{-8}$, seen in the broader red curve in Figure \ref{positions1a}.

In Figure \ref{positions1b} we present a histogram of the $y$-value at the point where the particles exit the computation box. The two tallest peaks correspond to the separatrices, with values between them corresponding to the reconnection outflow region. We see that the $\kappa = \eta_{sp}/\eta_a$ scattering model and the control case give very similar results, with 14\% and 12\% of the total particle orbits exiting within the outflow region respectively. In the case of much stronger scattering with $\kappa = 2\times 10^{-8}$ , significantly more orbits exit within the outflow region (20\% of total). Stronger scattering in the separatrices (in the case of the $\kappa = 2\times 10^{-8}$ case) causes more particle orbits to be scattered from the separatrices into the outflow region. Since no scattering takes place in this region and the $\textbf E\times \textbf B$ drift is directed outward, the test particles are unable to re-enter the current sheet and exit the simulation box in the outflow region. For higher values of $\kappa$, or for $\kappa = \eta_{sp}/\eta_a$, scattering is much weaker in the separatrices, resulting in a distribution of final $y$-values much closer to that of the control case.

\section{Discussion and conclusions}
\label{conclusions}

We have presented a very simple model of pitch angle scattering and have shown that it can have a significant impact on test particle energy spectra. In previous studies which included the effects of collisional scattering \citep[e.g.][]{numata-yoshida2002,burge-et-al2014}, it was found that repeated crossings of the reconnection region by test particles in the presence of scattering could lead to a higher energy gain than in the absence of scattering, but that the effect on energy spectra was not significant. In our work, the strong dependence of the mean free path on the anomalous resistivity is the main aspect of the model which affects the energy spectra and box escape positions. Due to this strong scattering, the spectra we obtain show a significant number of orbits gaining energies higher than is possible without scattering, which is something that is not seen in \citet{burge-et-al2014}, probably due to their use of a much lower scattering rate.

If we interpret $\kappa$ from Equation~\ref{mean-free-path} as the dependence of the mean free path on the anomalous resistivity, the difference between constant and spatially varying ($\kappa = \eta_{sp}/\eta_a$) values of $\kappa$ are mainly due to their behaviour in regions away from the central current sheet. Since our MHD simulations involved a constant anomalous resistivity where the current exceeded a specified threshold, whereas the Spitzer resistivity calculated at the location of the guiding centre is dependent on temperature, our choice of $\kappa = \eta_{sp}/\eta_a$ resulted in the scattering rate decreasing with temperature (this is a result of $\eta_{sp} \propto T^{-3/2}$). The weaker scattering in the separatrices due to the lower temperature (in comparison to the temperature inside the central current sheet) impacted the dynamics of the particles. Less scattering in the separatrices resulted in fewer orbits re-entering the current sheet multiple times. Therefore, the temperatures calculated in the MHD simulations have a significant effect on the test particle dynamics and energy spectra. The temperatures achieved in our MHD simulations are somewhat unrealistic, with a maximum temperature of $4.2\times 10^9 \unit K$, due to the lack of thermal conduction or radiation used. This resulted in the small values of $\eta_{sp}/\eta_a \approx 10^{-8}$ in the current sheet. In future work this could be remedied by the inclusion of thermal conduction and radiation in the MHD simulation. On the other hand, \citet{bian-et-al2016} showed that thermal conduction can be significantly reduced in coronal conditions due to pitch angle scattering, leading to temperatures of the order of $10^{8}\unit K$. The reduced thermal conductivity means it is reasonable that there is a large difference in temperature between the current sheet and the separatrices, resulting in a correspondingly large difference in the scattering rates and the associated test particle dynamics. We also showed that there are some small differences between an increasing and decreasing mean free path dependence as a function of the total test particle velocity, however they were negligible in comparison to the changes to the spectra as a result of varying $\kappa$.

The guiding centre formalism we used relies on the test particle being able to complete at least one full gyration in order to define a guiding centre, so the scattering rate is limited by the gyrofrequency. Our choice of scattering model, in particular $\kappa = \eta_{sp}/\eta_a$, would result in violating this restriction for values of anomalous resistivity more than an order of magnitude greater than the ones chosen. This may be remedied by solving full particle orbits for the time that the test particle is within the diffusion region \citep[as in][]{burge-et-al2014}. On the other hand, a further decrease in the anomalous resistivity in the MHD simulations would require a greater resolution, which would eventually take prohibitively long amounts of time to compute. 

In contrast to previous work on particle acceleration in 2D reconnection \citep[for instance in][]{gordovskyy-et-al2010a}, the maximum energies achieved in our simulations are relatively small (of the order of 100 \unit{keV}). This is a result of our use of a relatively small lengthscale causing small electric field strengths and size of reconnection region. For a given test particle orbit, the energy gain is entirely dependent on the electric potential drop that it traverses, with possible additional energy losses due to scattering. The presence of scattering introduces an additional lengthscale, namely the mean free path, which means that it is no longer possible to scale the resulting energy spectra with the MHD lengthscale. Our model of scattering did not include any energy loss during collisions, hence changes in the energy spectra are purely due to the different trajectories that particles take and the potential drop that they encounter along it. The inclusion of energy loss terms can be easily accommodated by adding stochastic terms to the energy evolution equation (Equation \ref{sde1}). This may result in an optimal anomalous resistivity for the acceleration of charged particles.

Finally, more complicated magnetic field topologies are likely to impact the results obtained in this paper with regards to particle trajectories and possibly energy spectra. It would be worth investigating how test particle acceleration is modified in 3D reconnection configurations such as that studied in \citet{threlfall-et-al2016a} or in coronal structures such as a flux tube \citep[see e.g.][]{gordovskyy-et-al2014} with the addition of anomalous scattering. 

\appendix

\section{Calculating $\dot\gamma$ and $\dot\beta$}
Since the guiding centre approach does not involve the total particle velocity, instead of the usual definition of the Lorentz factor we use,
\begin{align*}
\gamma^2 &= 1 + \frac{\gamma^2 v^2}{c^2} \\
&\simeq 1 + \frac{U^2}{c^2} + \frac{2\mu B}{mc^2} + \frac{\gamma^2 V_E^2}{c^2} \\
&= \frac{1 + \frac{U^2}{c^2} + \frac{2\mu B}{mc^2}}{1 - \frac{V_E^2}{c^2}},
\end{align*}
where we used the fact that the $\textbf E\times \textbf B$ is the dominant guiding centre drift. Therefore,
\begin{equation}
\gamma = \frac{\sqrt{1 + \frac{U^2}{c^2} + \frac{2\mu B}{mc^2}}}{\sqrt{1 - \frac{V_E^2}{c^2}}}. 
\end{equation}
Differentiating this with respect to time yields,
\begin{align}
\dot \gamma &= \frac 1 2 \left( 1 + \frac{U^2}{c^2} + \frac{2\mu B}{mc^2}\right)^{-1/2} \left( \frac{2U}{c^2}\frac{dU}{dt} + \frac{2\mu}{mc^2}\frac{dB}{dt} \right) \left( 1 - \frac{V_E^2}{c^2} \right)^{-1/2} \nonumber \\
&\hspace{1cm}+ \frac 1 2 \left( 1 - \frac{V_E^2}{c^2} \right)^{-3/2}\left( 1 + \frac{U^2}{c^2} + \frac{2\mu B}{mc^2}\right)^{1/2} \frac{2V_E}{c^2}\frac{dV_E}{dt}.
\end{align}
Since $V_E \ll c$, the second term is negligible in comparison to the first term, resulting in Equation \ref{sde3}.

For $\beta = \cos \theta$ the derivation of the time derivative, $\dot\beta,$ is much more straightforward. Since $\mu = \frac{mu^2 \beta^2}{2B} = \frac{mU^2}{2B} \frac{1-\beta^2}{\beta^2}$ and $\frac{d\mu}{dt} = 0$ we have:
\begin{align*}
0 = \frac{d\mu}{dt} &= \frac{mU}{B}\frac{1-\beta^2}{\beta^2}\frac{dU}{dt} + \frac{mU^2 \beta}{B}\left( -\frac{2}{\beta^3}\right) \frac{d\beta}{dt} - \frac{m U^2}{2B^2} \frac{1-\beta^2}{\beta^2}\frac{dB}{dt} \\
&= \frac{2\mu}{U}\frac{dU}{dt} - \frac{2\mu}{\beta\left( 1-\beta^2 \right)}\frac{d\beta}{dt} - \frac{\mu}{B} \frac{dB}{dt}.
\end{align*}
Therefore, the time derivative, $\dot \beta$, can be expressed as:
\begin{equation}
\dot \beta = \left( \frac{1}{U}\frac{dU}{dt} - \frac{1}{2B}\frac{dB}{dt} \right) \beta\left( 1-\beta^2 \right).
\end{equation}

\begin{acknowledgements}
A.B. would like to thank the University of St Andrews for financial support from the 7th Century Scholarship and the Scottish Government for support from the Saltire Scholarship. E.P.K.'s work is partially supported by a STFC consolidated grant ST/L000741/1. J.T. and T.N. gratefully acknowledge the support of the UK STFC (consolidated grant SN/N000609/1).
\end{acknowledgements}

\bibliographystyle{aa}
\bibliography{$HOME/Desktop/papers/references}

\end{document}